%% file: main.tex
\documentclass[sigconf,screen,nonacm]{acmart}

\AtBeginDocument{%
  }

\newif\ifappendix
\appendixtrue

\usepackage{tikz}
\usepackage{amsmath}

\usepackage{filecontents}

\usepackage{amsmath}
\usepackage{amsthm}
\usepackage{amssymb}
\usepackage{mathtools}
\usepackage{bm}
\usepackage{bbm}
\usepackage{booktabs}
\usepackage{xspace}
\usepackage{color,soul}
\usepackage{multirow}
\usepackage{multicol}
\usepackage{caption}
\usepackage{enumitem}
\usepackage{verbatim}
\usepackage{microtype}
\usepackage{graphicx}
\usepackage{booktabs}
\usepackage{hyperref}
\usepackage{nicefrac}
\usepackage[normalem]{ulem}
\usepackage{tikz}
\usetikzlibrary{trees}
\usepackage{float}
\usepackage{subcaption}
\usepackage[ruled,noend,linesnumbered,vlined]{algorithm2e}
\usepackage[dvipsnames]{xcolor}
\usepackage{colortbl}
\usepackage{hhline}
\usepackage{tabularx}
\usepackage{listings}
\usepackage{minted}

\usepackage{listings}
\newcommand{\myvspace}[1]{\vspace{#1}}

\definecolor{myred}{rgb}{1.0,0.7,0.8}
\definecolor{mygreen}{RGB}{0,166,0}
\definecolor{lightgreen}{rgb}{0.56, 0.93, 0.56}
\definecolor{myorange}{RGB}{252,107,4}
\definecolor{darkgreen}{RGB}{0,153,102}
\definecolor{lightblue}{rgb}{0.53, 0.81, 0.92}
\definecolor{lightgray}{gray}{0.9}

\newcommand{\mysubsubsection}[1]{\subsubsection*{\textbf{\ul{#1}}}}

\DeclareMathOperator*{\argmin}{arg\,min}

\newcommand{\system}{{\text{Hydraulis}}\xspace}

\newtheorem{problem}{Problem}

\newtheorem*{assumption*}{\assumptionnumber}
\providecommand{\assumptionnumber}{}


\newtheorem*{statement*}{\statementnumber}
\providecommand{\statementnumber}{}


\begin{document}

\title{\system: Balancing Large Transformer Model Training via Co-designing Parallel Strategies and Data Assignment}
\renewcommand{\shorttitle}{\system}

\author{Haoyang Li}
\authornote{School of Computer Science \& Key Lab of High Confidence Software Technologies (MOE), Peking University}
\email{lihaoyang@stu.pku.edu.cn}
\affiliation{
\institution{Peking University}
\country{China}
}

\author{Fangcheng Fu}
\authornote{School of Artificial Intelligence, Shanghai Jiao Tong University}
\email{ccchengff@sjtu.edu.cn}
\affiliation{
\institution{Shanghai Jiao Tong University}
\country{China}
}

\author{Sheng Lin}
\authornotemark[1]
\email{linsh@stu.pku.edu.cn}
\affiliation{
\institution{Peking University}
\country{China}
}

\author{Hao Ge}
\authornotemark[1]
\email{gehao@stu.pku.edu.cn}
\affiliation{
\institution{Peking University}
\country{China}
}

\author{Xuanyu Wang}
\authornotemark[1]
\email{wxyz0001@pku.edu.cn}
\affiliation{
\institution{Peking University}
\country{China}
}

\author{Jiawen Niu}
\authornotemark[1]
\email{niujiawen705@stu.pku.edu.cn}
\affiliation{
\institution{Peking University}
\country{China}
}

\author{Jinbao Xue}
\email{jinbaoxue@tencent.com}
\affiliation{
\institution{Tencent}
\country{China}
}

\author{Yangyu Tao}
\email{brucetao@tencent.com}
\affiliation{
\institution{Tencent}
\country{China}
}

\author{Di Wang}
\email{diwang@tencent.com}
\affiliation{
\institution{Tencent}
\country{China}
}

\author{Jie Jiang}
\email{zeus@tencent.com}
\affiliation{
\institution{Tencent}
\country{China}
}

\author{Bin Cui}
\authornotemark[1]
\authornote{Institute of Computational Social Science, Peking University (Qingdao)}
\email{bin.cui@pku.edu.cn}
\affiliation{
\institution{Peking University}
\country{China}
}

\renewcommand{\shortauthors}{Haoyang Li et al.}

\begin{abstract}
% To optimize large Transformer model training, efficient parallel computing and advanced data management are essential.
To optimize large Transformer model training, both efficient parallel computing and advanced data management are indispensable. % However, current methods often assume a stable and uniform training workload, neglecting imbalances in data sampling and packing that can impede performance. 
However, current methods often assume a stable and uniform training workload, neglecting data-induced imbalances—arising from both sampling and packing processes—which can impede training performance.
Specifically, data sampling imbalance arises from uneven sequence length distribution of the training data, while data packing imbalance stems from the discrepancy between the linear memory complexity and quadratic time complexity of the attention mechanism. 
To address these imbalance issues, we develop \system, which jointly optimizes the parallel strategies and data assignment. 
For one thing, we introduce large model training with dynamic heterogeneous parallel strategies in response to the sequence length variations within and across training iterations. 
For another, we devise a two-stage data assignment approach, which strikes a good balance in terms of the training workloads both within and across model replicas. 
Empirical results demonstrate that \system outperforms existing systems by 1.32-2.66$\times$. 
Our source code is available: 
{\url{https://github.com/PKU-DAIR/Hetu}}.
\end{abstract}

\maketitle
\pagestyle{plain}

\input{sections/intro}

\input{sections/prelim}

\input{sections/analysis}
\input{sections/method}
\input{sections/exp}

\input{sections/ending}

%-------------------------------------------------------------------------------

\begin{acks}
This work is supported by National Natural Science Foundation of China (U23B2048, 62402011), PKU-Tencent joint research Lab, and High-performance Computing Platform of Peking University. 
Fangcheng Fu and Bin Cui are the corresponding authors.
\end{acks}

\bibliographystyle{ACM-Reference-Format}
\bibliography{reference}

\clearpage
\appendix
\input{sections/appendix}

\end{document}

%% file: sections/intro.tex
\section{Introduction}
\label{sec:intro}

In recent years, large Transformer models~\cite{vaswani2017attention} have become pivotal in fields such as natural language processing~\cite{devlin2019bert, brown2020language, raffel2020exploring}, computer vision~\cite{carion2020end, dosovitskiy2021image, liu2021swin}, video processing~\cite{bertasius2021space, arnab2021vivit, fan2021multiscale}, speech recognition~\cite{dong2018speech, baevski2020wav2vec, gulati2020conformer} and intelligent agents~\cite{db-gpt, dl-code, llm-agents, agent-rise}. Their success relies on pre-training models with vast parameters on extensive datasets~\cite{kaplan2020scaling, dubey2024llama3}.

\begin{figure}[!t]
    \centering
    \includegraphics[width=\linewidth]{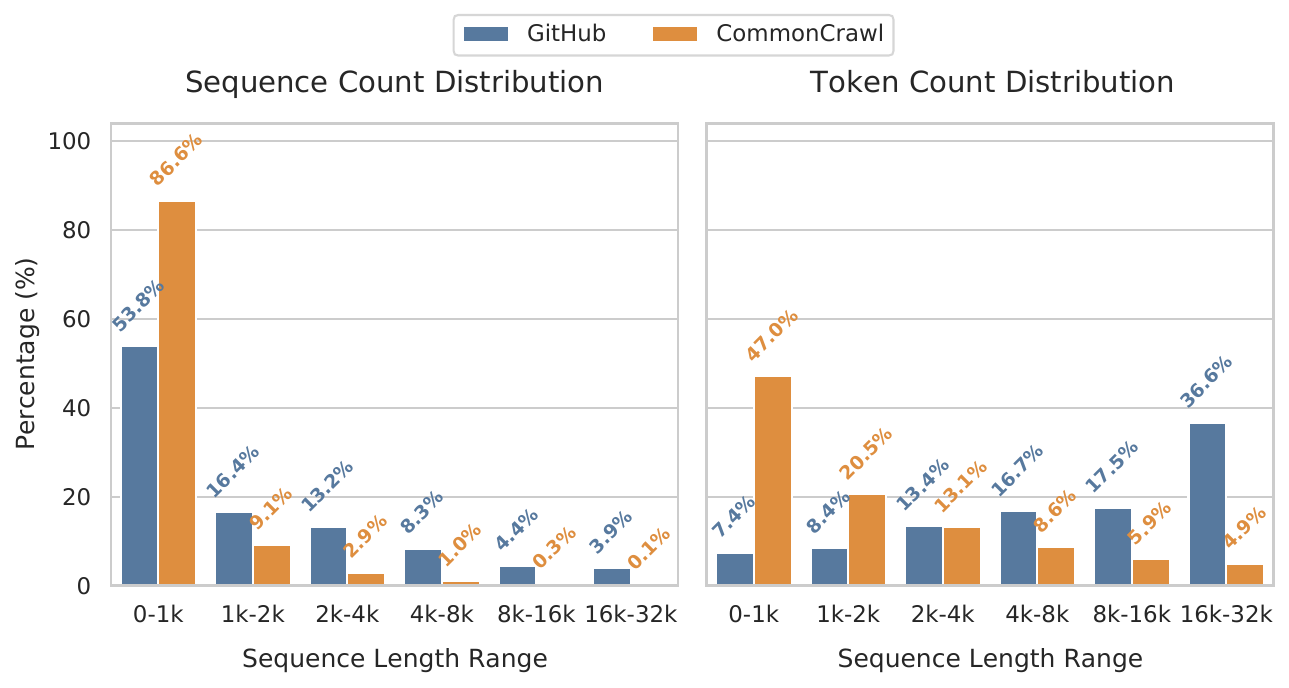}
    \caption{\small{Sequence length distribution of two popular open-sourced datasets. We display the number of sequences and tokens within each length range, with both datasets exhibiting high variance.}}
    \label{fig:distribution}
    \myvspace{-10pt}
\end{figure}

Within this landscape, two key challenges arise. (a) To achieve better model ability, efforts have been made to scale models to larger sizes~\cite{kaplan2020scaling}, making large model training substantially time-consuming and memory-intensive.
(b) Meanwhile, to move towards Artificial General Intelligence (AGI), input data are often condensed into sequences of tokens to achieve a unified representation~\cite{mikolov2012subword,dosovitskiy2021image,llm-dataset}. However, this representation naturally exhibits length variability~\cite{seq_length_of_corpora}. For instance, Figure~\ref{fig:distribution} presents the sequence length distribution after tokenizing two popular large language models (LLMs) training datasets—CommonCrawl and GitHub—revealing that the sequence lengths are significantly diverse in the real world. Thus, large model training takes as input variable-length data.

To address this, approaches from two perspectives have been developed. 
(a) From a parallel computing perspective, methods like data parallel~\cite{li2020pytorch, you2019large, zhang2024simplefsdp, bai2022modern}, tensor parallel~\cite{shoeybi2019megatron, narayanan2021efficient, korthikanti2022reducing}, pipeline parallel~\cite{huang2019gpipe, kim2020torchgpipe, narayanan2021memory, lamypoirier2023breadthfirst, qi2023zerobubble, harlap2018pipedream, pipeline-overview}, sequence and context parallel~\cite{jacobs2023deepspeedulysses, liu2023ring, li2024distflashattn, brandon2023stripedattention, gu2024loongtrain, fang2024usp, flexsp} have been proposed, and are often combined to train large-scale models.
Data parallel operates at the granularity of training pipelines (i.e., data parallel groups containing model replicas that are capable of independent forward and backward propagation) and thus can typically scale by adding more pipelines, while other parallel methods (i.e., tensor, pipeline, sequence and context parallel) are optimized within a single training pipeline to balance pipeline latency and memory usage.
(b) From a data management perspective, data packing techniques—combining shorter sequences into longer ones for simultaneous processing—are widely used in the training of both language~\cite{krell2022packing, nvidia2023packing, bai2024longalign, wang2024packing, kundu2024packing} and vision~\cite{dehghani2023patchnpacknavit, choudhury2024dontlooktwicefaster} models to minimize redundant computations.

Despite these advancements, existing training systems typically leverage these two perspectives in a decoupled manner. 
(a) In parallel computing, most systems overlook the sequence length variation and adopt a static, homogeneous parallel strategy~\cite{shoeybi2019megatron, rasley2020deepspeed, zheng2022alpa, jia2019beyond}, which deploys all training pipelines (a.k.a. model replicas) with the same parallel scheme throughout the training (e.g., all pipelines have identical tensor parallel and pipeline parallel degree).\footnote{
In this work, we use ``parallel scheme'' to denote the parallel configuration of a pipeline (each pipeline does not necessarily employ any parallel methods, i.e., it can be deployed on only one GPU).
Given the available GPUs, a ``parallel strategy'' describes the collection of parallel schemes of all pipelines deployed.}
(b) In data management, existing works usually specify a maximum sequence length (a.k.a. context length), pack the original sequences into multiple packed ones that do not exceed the maximum length (termed max-length packing), and distribute the packed sequences evenly across machines~\cite{nvidia2023packing, kundu2024packing}.

In short, current works enforce a static, homogeneous parallel strategy and a max-length packing on variable-length data. 
However, our empirical analysis (details shown in \S\ref{sec:analysis}) reveals that this decoupled design overlooks two critical imbalances.

Firstly, sequences in the real world substantially vary in length by nature~\cite{seq_length_of_corpora}. 
This leads to the \textit{\ul{data sampling imbalance}}, manifesting as variable sequence lengths within iterations and inconsistent maximum lengths across iterations.
When employing a homogeneous parallel strategy, we must ensure the parallel scheme is memory-saving enough (e.g., with a high context parallel or tensor parallel degree) to support the maximum sequence length, which compromises the efficiency of short sequences.

The second is the \textit{\ul{data packing imbalance}}. 
Due to the attention mechanism in Transformer, there is a discrepancy between the quadratic time complexity and the linear memory complexity w.r.t. the sequence length~\cite{vaswani2017attention, dao2022flashattention, nvidia2023packing}. 
However, existing works adopt max-length packing, achieving memory balance but leaving the training workloads imbalanced within and across pipelines. 

Motivated by this, we manage to address these imbalance issues by jointly optimizing the parallel and packing strategies, putting forward co-designs in both the parallel computing and data management perspectives. 
To this end, we introduce \system, an innovative system designed for efficient training of large Transformer models. Our major contributions are outlined as follows.

\textbf{(\uppercase\expandafter{\romannumeral1}) Training with dynamic heterogeneous parallel strategies.} 
Unlike conventional training that relies on a static, homogeneous parallel strategy, we employ dynamic, heterogeneous strategies to address the \textit{\ul{data sampling imbalance}}. Each heterogeneous strategy can deploy pipelines with varying parallel schemes to handle sequences of varying lengths. Divergent strategies can be used for different iterations, allowing each iteration to employ the most suitable strategy for the ad hoc sequence length distribution. 

\textbf{(\uppercase\expandafter{\romannumeral2}) Disaggregating optimization and propagation with subgraphs.} 
To support training with dynamic heterogeneous parallel strategies, we present a novel disaggregation of the optimization phase (i.e., synchronizing gradients and updating the model) and propagation phase (i.e., executing forward and backward propagation in a pipelined manner) in distributed training.
This approach maintains consistent model states sharding for optimization, and can transit to arbitrary heterogeneous parallel strategy for propagation.
We further introduce subgraphs (i.e., subsets of the complete computation graph) to unify the representation of sophisticated communication needs during propagation (for activations) and optimization-propagation interactions (for parameters and gradients). 
This allows for flexible combinations of parallel schemes and seamless transitions between parallel strategies.

\textbf{(\uppercase\expandafter{\romannumeral3}) Two-stage sequence assignment.} 
Given a mini-batch of sequences and an arbitrary strategy, we devise a two-stage approach to eliminate the \textit{\ul{data packing imbalance}}. 
In particular, we formulate an optimization problem that aims to determine how packed sequences can be balanced within each pipeline.
Based on this, we also strike a good balance across different pipelines by carefully dispatching the original (unpacked) sequences.
Besides, this process also estimates latency for each possible strategy, allowing us to evaluate and select the best one for each iteration.

\textbf{(\uppercase\expandafter{\romannumeral4}) Data distribution-aware strategy proposal.} 
Guided by the sequence assignment and for practical considerations, we further propose preparing the candidate strategies that are promising in advance, in order to accelerate the strategy enumeration and selection for each iteration. 
It takes the overall sequence length distribution of the entire dataset into account, and proposes the candidate strategies via a dynamic programming process. 

\textbf{(\uppercase\expandafter{\romannumeral5}) Implementation and Evaluation.} 
We implement \system on top of Hetu\footnote{\url{https://github.com/PKU-DAIR/Hetu}}~\cite{hetu-v2} and conduct extensive experiments to demonstrate that \system achieves 1.32-2.66$\times$ improvement over state-of-the-art training systems, while also offering better scalability.

%% file: sections/prelim.tex
\section{Preliminaries}
\label{sec:prelim}

\subsection{Data Sampling and Data Packing}
\label{subsec:data-sampling}

During the training of a Transformer model, each iteration samples a mini-batch of sequences with a pre-defined threshold on the total number of tokens, performs forward and backward propagation on the mini-batch, and updates the model. 
 
Besides, the model is typically trained with a pre-defined context length, representing the maximum number of tokens it can process in a single input~\cite{pawar2024contextlength}. Any sequence exceeding this limit is typically truncated to fit within the context length. 
But regardless of whether the original sequences are truncated or not, the lengths of the sampled sequences will inevitably exhibit variance. 

This variability introduces challenges when constructing batches, as sequences need to be aligned to a common length for efficient batch processing. Traditionally, this issue is addressed through data padding, where shorter sequences are padded with special tokens (e.g., [PAD] tokens) to match the length of the longest one~\cite{devlin2019bert, vaswani2017attention}. Although padding ensures uniformity in sequence length and is straightforward to implement, it leads to unnecessary computation on the padded tokens, resulting in wasted computational resources.

\begin{figure}[!t]
    \centering
    \includegraphics[width=\linewidth]{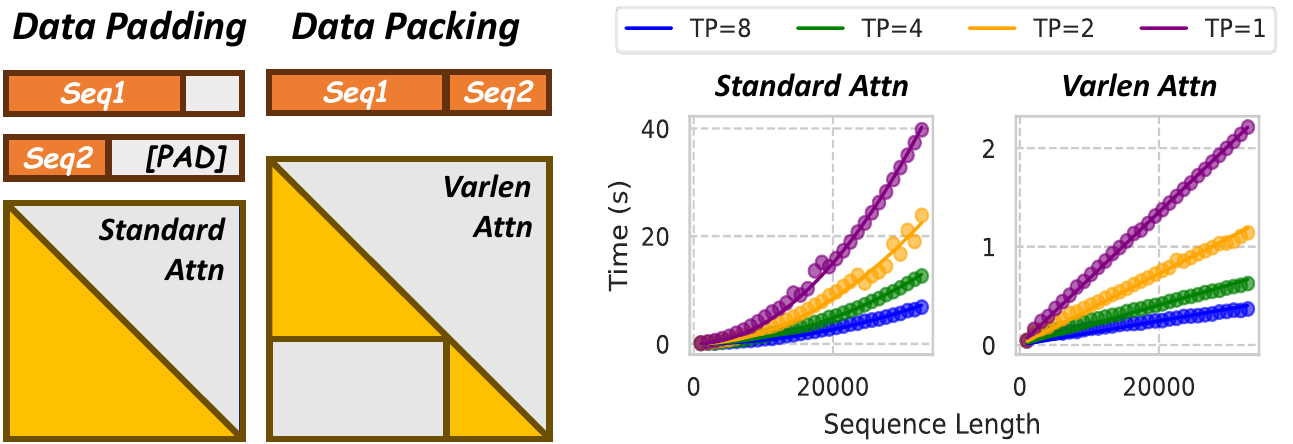}
    \caption{\small{Left: Illustration of padding and packing. Right: Comparison of the attention latency during LLaMA2 7B training on Nvidia A800 GPUs with tensor parallel (TP). 
    % Note that varlen attention uses 1K as a packing unit. 
    Standard attention scales quadratically, while varlen achieves linear scaling relative to the sequence length (1K) before packing.}}
    \myvspace{-10pt}
    \label{fig:padding_and_packing}
\end{figure}

\begin{figure*}[!t]
    \begin{subfigure}{0.4\linewidth}
        \centering
        \includegraphics[width=\linewidth]{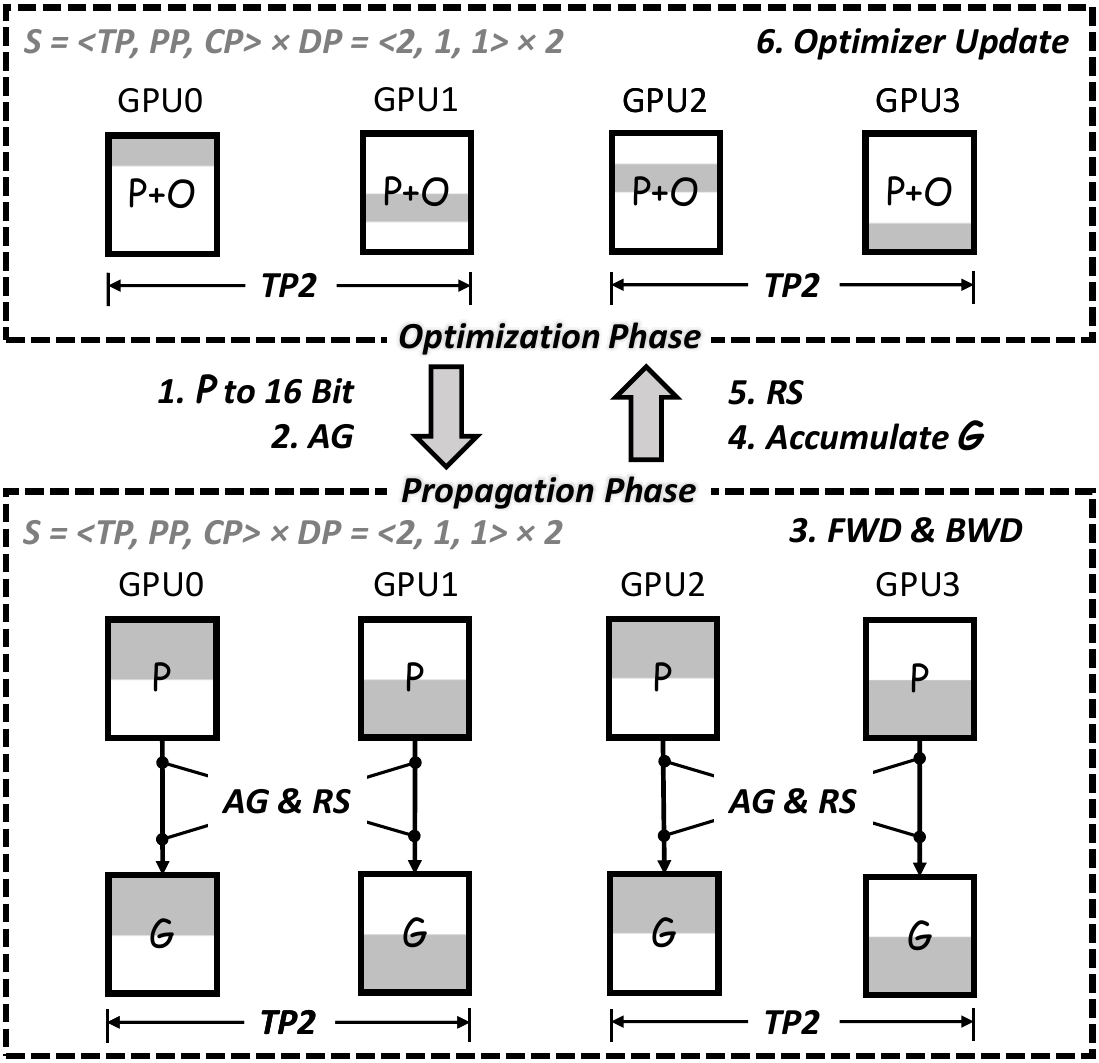}
        \subcaption{Standard training workflow.}
        \label{subfig:standard_op}
    \end{subfigure}
    \begin{subfigure}{0.4\linewidth}
        \centering
        \includegraphics[width=\linewidth]{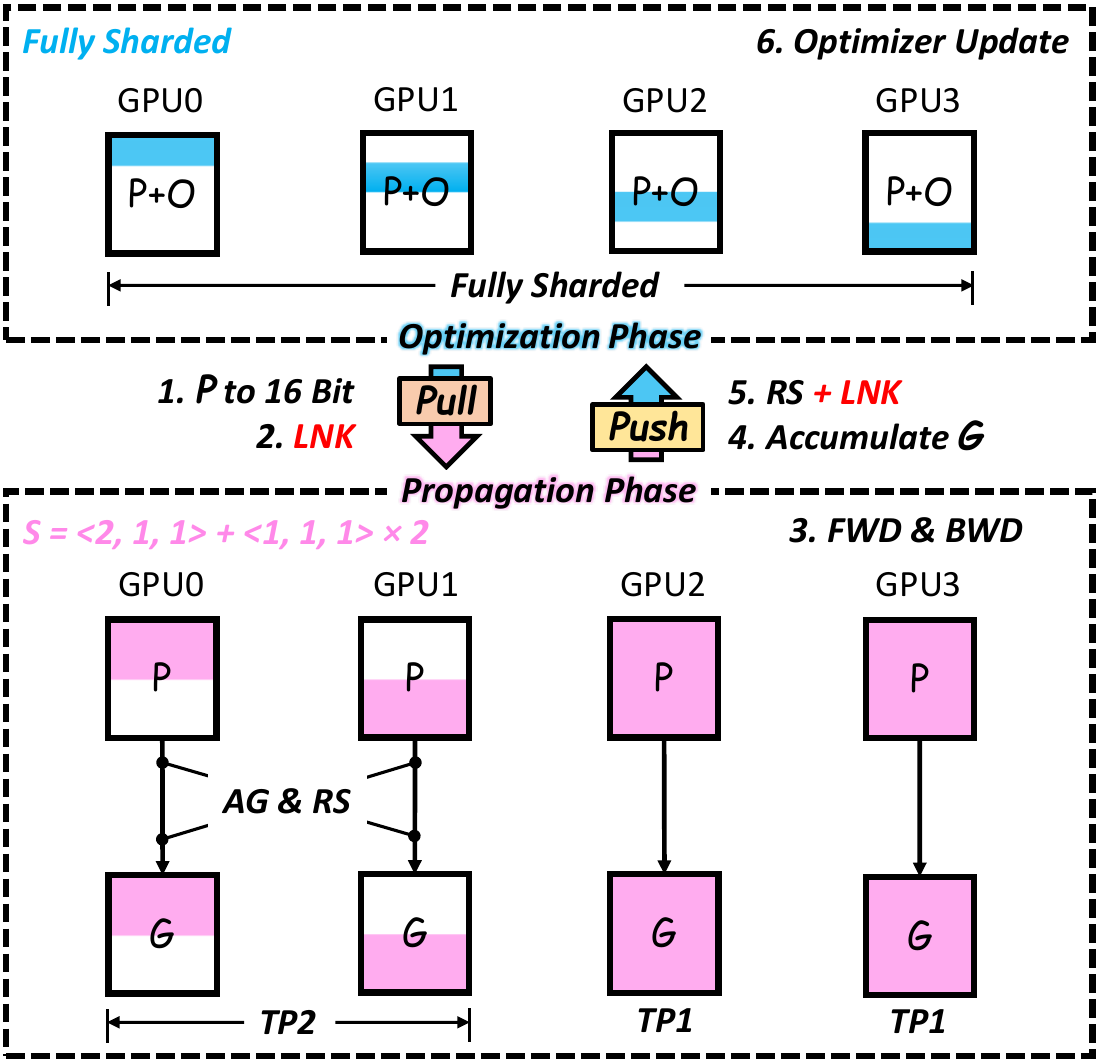}
        \subcaption{Disaggregate optimization and propagation.}
        \label{subfig:disaggregate_op}
    \end{subfigure}
    \caption{\small{``O, P, G'' represents optimizer states, parameters and gradients, respectively. ``AG, RS, S, R'' represents all-gather, reduce-scatter, send and receive, respectively. ``To'' represents the data type transfer operator (from 32- to 16-bit floating points), and ``LNK'' represents the link operator designed by us. (a) The overall workflow of the model states sharding technique combined with mixed-precision training. (b) Our proposed optimization-propagation disaggregation technique, which decouples the distributed strategies for the optimization phase and propagation phase, allowing for arbitrary heterogeneous parallel strategies during propagation.}}
    \label{fig:op}
    \myvspace{-10pt}
\end{figure*}

As shown in Figure~\ref{fig:padding_and_packing}, an alternative to padding is packing~\cite{krell2022packing}, which seeks to maximize computational efficiency by grouping sequences of varying lengths into a single longer sequence, thereby eliminating the need for padded tokens. To prevent cross-contamination between the packed sequences, positional embeddings and the attention mask must be adjusted to preserve the integrity of the attention mechanism~\cite{bai2024longalign, wang2024packing, kundu2024packing}. Fortunately, modern techniques such as FlashAttention v2 have integrated support for block diagonal attention, which natively handles variable-length sequences (i.e., varlen)~\cite{dao2022flashattention, dao2023flashattention2}. Moreover, the varlen interface ensures that the attention complexity scales as $\sum{S_{i}}^2$, rather than $(\sum{S_{i}})^2$, where $S_{i}$ represents the length of the $i$-th original sequence being packed~\cite{nvidia2023packing}. This significantly reduces the computational overhead of data packing, making it a more practical and efficient approach.

\subsection{Parallel Methods, Schemes, and Strategies}
\label{subsec:preliminary_parallel}

As models grow in both parameter size and computational complexity, distributed training across multiple devices becomes essential. 
Below we introduce popular parallel methods for large Transformer model training. 

\mysubsubsection{Data parallel}
In data parallel (DP), the model is replicated across multiple devices, with each device processing a different subset of the input data~\cite{li2020pytorch, you2019large, zhang2024simplefsdp, bai2022modern}. After computing the gradients locally, devices must synchronize their model updates, typically through a gradient reduction process, to ensure consistency across devices.

\mysubsubsection{Parallel scheme of one pipeline}
In this work, we call each data parallel group one training pipeline, and the parallel scheme describes how a pipeline is deployed. 
We can broadly classify parallel methods for a pipeline into inter- and intra-op parallel, respectively.

The Inter-op parallel is mainly implemented with pipeline parallel (PP).  
Particularly, PP divides the model into sequential stages, with each stage assigned to a different device. Data from a mini-batch is further split into micro-batches, which flow through the pipeline in a staged manner. 
(When training Transformer models, each micro-batch accounts for one packed sequence.)
Techniques such as micro-batch scheduling and gradient accumulation are often employed, allowing multiple micro-batches to be processed efficiently~\cite{huang2019gpipe, kim2020torchgpipe, narayanan2021memory, lamypoirier2023breadthfirst, qi2023zerobubble, harlap2018pipedream, pipeline-overview}.

Unlike inter-op parallel, there are several intra-op parallel methods for a pipeline. 
Tensor parallel (TP) partitions each layer across multiple GPUs and exchanges intermediate results (e.g., via all-gather and reduce-scatter operations) to complete the forward and backward passes~\cite{shoeybi2019megatron, narayanan2021efficient, korthikanti2022reducing}. 
Additionally, sequence parallel (SP) and context parallel (CP) have been explored to handle extremely long sequences, where a single sequence is partitioned across multiple devices to avoid out-of-memory (OOM) errors, and all-to-all or ring-style peer-to-peer communication would be necessary to perform the attention mechanism over the partitioned sequences~\cite{jacobs2023deepspeedulysses, liu2023ring, li2024distflashattn, brandon2023stripedattention, gu2024loongtrain, fang2024usp}.

\mysubsubsection{Parallel strategy}
Building on the parallel scheme of a single pipeline, the parallel strategy refers to the collection of parallel schemes of all pipelines. % deployed over all available GPUs. 
However, existing works mainly leverage the same parallel schemes for all pipelines, resulting in a homogeneous parallel strategy. 
Such a homogeneous design fails to take account of the imbalances caused by data sampling and data packing, which will be discussed in \S\ref{sec:analysis}.

\subsection{Model States Sharding}
\label{subsec:model-states-sharding}

To reduce memory consumption during training, model states sharding techniques (a.k.a. ZeRO~\cite{rajbhandari2020zero} and FSDP~\cite{zhao2023fsdp, zhang2024simplefsdp}) are widely used. These methods extend standard data parallel (DP) by partitioning model states (i.e., parameters, gradients, and optimizer states) across multiple devices, rather than replicating them across all data parallel groups (i.e., pipelines).

The sharding technique is combined with mixed-precision training~\cite{micikevicius2018mixedprecision, nvidia2023mixedprecision}, which uses both 16-bit floating point (FP16 or BF16) and 32-bit floating point (FP32) precisions to balance efficiency and numerical stability. 
Figure~\ref{subfig:standard_op} illustrates an example with each pipeline deployed with a tensor parallel degree of 2. 

The 32-bit optimizer states and parameters (which consume more memory but are critical for ensuring convergence) are sharded across GPUs, 
while the 16-bit parameters and gradients (which are more memory-efficient and computationally faster) are maintained during propagation.

As aforementioned, current works mainly adopt a homogeneous parallel strategy, so they usually align the sharding with the parallel strategy, allowing distributed state transformations to be easily achieved via all-gather and reduce-scatter operations. 

However, since such transformations can overlap with computation, it is also feasible to adopt more nuanced, non-aligned parameter pulling and gradient pushing using customized communication strategies, which is depicted in Figure~\ref{subfig:disaggregate_op} and will be elaborated in \S\ref{subsec:op_disaggregation}.

\subsection{Graph-based Representations}
\label{subsec:graph}

The execution of a deep learning model can be represented as a computation graph—a directed acyclic graph (DAG) where nodes correspond to operators (e.g., matrix multiplication, attention) defined by users and edges indicate data dependencies. Modern deep learning frameworks, such as PyTorch~\cite{li2020pytorch} and TensorFlow~\cite{abadi2016tensorflow}, adopt this graph-based representation to encapsulate execution logic. Following this convention, we employ the same representation but extend it with a subgraph abstraction to enable dynamic heterogeneous strategies, which will be presented in \S\ref{subsec:subgraphs}.

\subsection{Limitations of State-of-the-art Systems}
State-of-the-art (SOTA) training systems, including Megatron-LM~\cite{shoeybi2019megatron, narayanan2021efficient, korthikanti2022reducing} and DeepSpeed~\cite{rasley2020deepspeed, jacobs2023deepspeedulysses}, have incorporated diverse parallel methods (e.g, Megatron-LM supports DP, TP, PP and CP, while DeepSpeed offers DP and SP) and model states sharding techniques (e.g., Megatron-LM implements ZeRO-1, which shards the optimizer states and the FP32 parameters, while DeepSpeed supports ZeRO 1-3 to allow further sharding FP16/BF16 gradients and parameters).
However, they are limited to using one static parallel strategy throughout the training task. A more recent work, HotSPa~\cite{ge2023hotspa}, enables the use of multiple pre-defined parallel strategies and supports switching between them within an iteration. This enables data in different sequence lengths within an iteration to utilize different parallel strategies while accumulating gradients for a global update.

Nonetheless, as will be analyzed in \S\ref{sec:analysis}, various aspects of data imbalance necessitate support for dynamic heterogeneous strategies. However, all of these systems lack this capability, as they are constrained to homogeneous parallel strategies (i.e., all pipelines use the same parallel scheme) and require model states sharding to align with the parallel strategy. This is due to the following reasons:
(a) Existing designs do not support parameter retrieval and gradient aggregation across heterogeneous pipelines, as they rely on homogeneous communication primitives (i.e., all-gather and reduce-scatter) but lack heterogeneous ones.
(b) Balancing workload across heterogeneous pipelines is challenging due to the absence of fine-grained data assignment mechanisms and strategy design methods that take data distribution into account.

To address (a), \system adopts an optimization-propagation disaggregation architecture to facilitate sophisticated parameter retrieval and gradient aggregation, and introduces subgraphs to support arbitrary heterogeneous strategies (\S\ref{sec:system_designs}). To address (b), it proposes a two-stage sequence assignment method for load balancing (\S\ref{sec:sequence_management}), and pre-generates distribution-aware heterogeneous parallel strategies before training (\S\ref{sec:strategies_generator}). Together, these innovations overcome the limitations of current training systems and improve training efficiency under data imbalance conditions (\S\ref{sec:exp}).

%% file: sections/analysis.tex
% \myvspace{-8pt}
\section{Imbalances Analysis}
\label{sec:analysis}

In this section, we analyze the imbalance issues encountered during data sampling and packing in training on real-world datasets, and discuss potential optimization opportunities.
To ease the analysis, we exemplify the imbalance issues by training the LLaMA 13B model with the CommonCrawl~\cite{tiiuae2024falconrefinedweb} and GitHub~\cite{codeparrot2024githubcode} datasets over 16 GPUs, with 32K context length and 100K tokens per iteration.

\begin{figure}[!t]
\centering
\includegraphics[width=\linewidth]{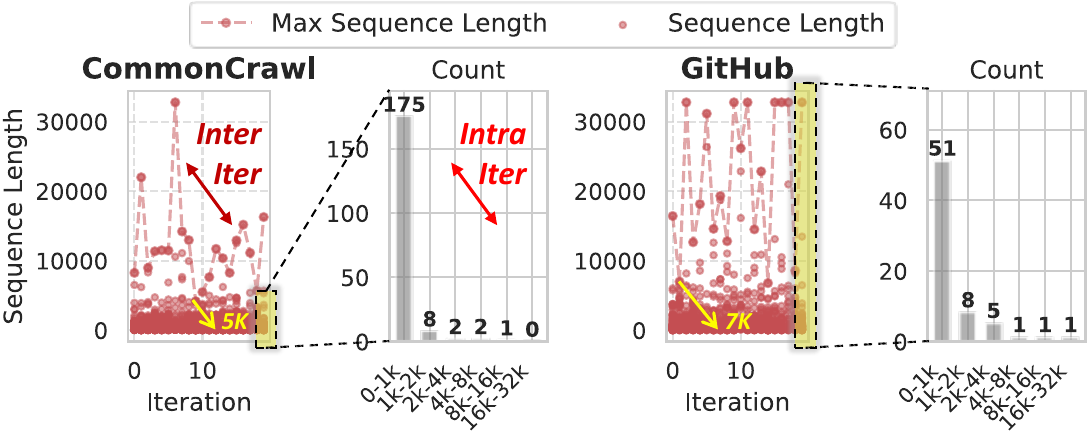}
\caption{\small{
The dot plot on the left shows the fluctuation of maximum sequence length, emphasizing inter-iteration imbalance, while the bar chart on the right presents the distribution of sequence lengths within an iteration (20th iteration), highlighting intra-iteration imbalance.}}
\label{fig:data_sampling_imbalance}
\myvspace{-10pt}
\end{figure}

\subsection{Data Sampling Imbalance}

We first concentrate on the imbalance in terms of data sampling.
As shown in Figure~\ref{fig:data_sampling_imbalance}, we observe two key issues: (a) within iterations, sequence lengths are unevenly distributed, with more short sequences and fewer long ones; (b) across iterations, the sequence distribution varies significantly, and the maximum sequence length fluctuates considerably, often falling short of the context length. In the following, we analyze the two issues, respectively.

\mysubsubsection{Intra-iteration imbalance}
The uneven sequence length distribution within an iteration can be attributed to the inherent property of data. In particular, sequences in the real world generally follow a long-tail distribution, where short sequences dominate and long sequences are rare. Existing studies have also observed such a distribution in many popular training datasets~\cite{seq_length_of_corpora}. 
This imbalance poses both challenges and opportunities when it comes to optimizing overall performance. 

\begin{figure}[!t]
\centering
\includegraphics[width=\linewidth]{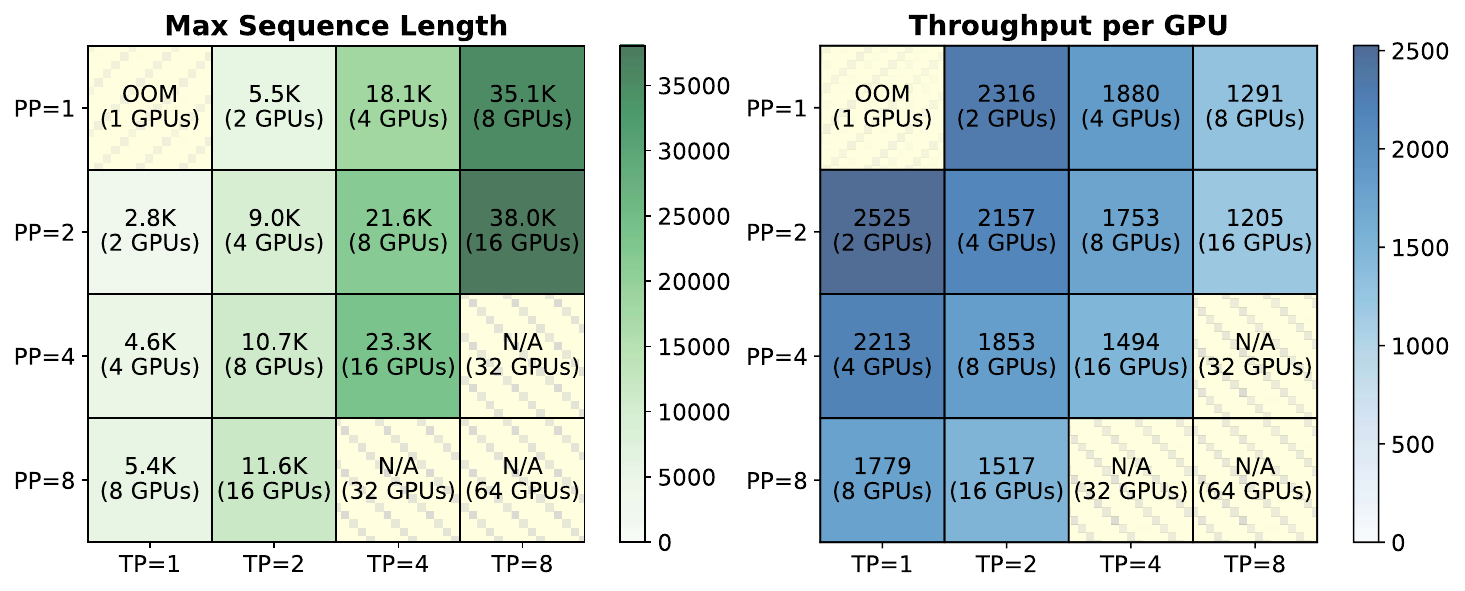}
\caption{\small{Memory and throughput trade-offs. ``OOM'' indicates out-of-memory, and ``N/A'' indicates not available. The experiment is conducted using a 13B LLaMA model with 16 Nvidia A800 GPUs. Throughput is measured in tokens per second and maximum sequence lengths are given by $\texttt{MaxLen}(\cdot)$ (see \S\ref{subsec:sequence_packing} and \ifappendix{Appendix~\ref{appendix:memory_model}}\else{Appendix C.1}\fi~\cite{hydraulis_appendix}).}}
\label{fig:memory_vs_throughput}
\myvspace{-15pt}
\end{figure}

The challenges stem from the trade-offs between memory usage and throughput offered by different numbers of GPUs and parallel schemes. As shown in Figure~\ref{fig:memory_vs_throughput}, increasing the number of GPUs generally leads to a lower throughput per GPU but allows for a longer maximum sequence length. Meanwhile, under the same number of GPUs, different parallel schemes (e.g., PP, TP and their combinations) also have divergent trade-offs (e.g., replacing PP with TP generally supports longer sequences but at the cost of reduced per-GPU throughput). 
As aforementioned in \S\ref{subsec:preliminary_parallel}, existing works utilize all GPUs in the cluster uniformly with a homogeneous parallel strategy tailored for the longest context length, compromising the overall training efficiency for memory.

Nevertheless, combined with the length imbalance within a mini-batch, such trade-offs become particularly invaluable:
short sequences can be processed using faster but more memory-intensive parallel schemes, while the long sequences can be assigned to slower yet more memory-efficient parallel schemes. 
Therefore, adopting heterogeneous parallel strategies--- where non-unique parallel schemes are applied on different proportions of GPUs based on different sequence lengths--- enables better performance (\S\ref{sec:system_designs}).

\mysubsubsection{Inter-iteration imbalance}

While intra-iteration imbalance arises from the distribution of sequence lengths within a single mini-batch, another challenge comes from the distribution variations across mini-batches. 
Particularly, due to the long-tail distribution, rare long sequences may not appear in every iteration (even with a large batch size on a 1024-GPU cluster, the maximum sequence length still varies, as detailed in \ifappendix{Appendix~\ref{appendix:larger_batch_size}}\else{Appendix A.1}\fi~\cite{hydraulis_appendix}). 
This leads to fluctuations in data distribution and maximum sequence lengths across iterations. 
However, current works overlook this issue, using a static parallel strategy for all iterations, which is unsuitable. 
For example, when an iteration consists mainly of short sequences, parallel schemes designed for long sequences become overkill.
To address this, we propose to dynamically adjust the parallel strategy based on the characteristics of each mini-batch to better align the workload, thereby enhancing overall performance across iterations (\S\ref{sec:system_designs}).

\begin{figure}[!t]
    \centering
    \includegraphics[width=\linewidth]{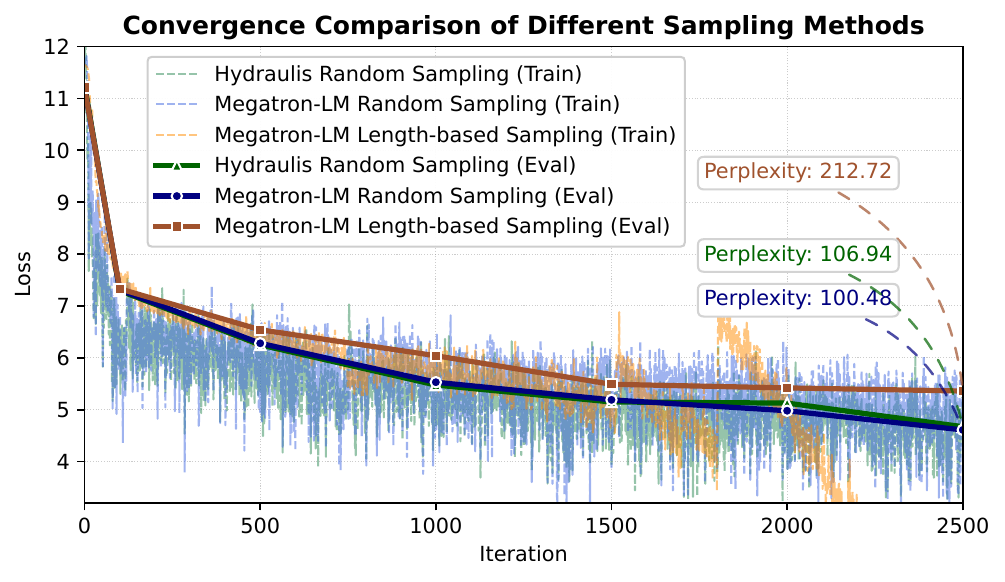}
    \caption{\small{Training and evaluation loss of different sampling methods. We subsample 250M tokens for training and 5M for validation from the GitHub dataset, and train a 13B model on 16 GPUs with 32K context length and 100K tokens per iteration.}}
    \label{fig:loss}
    \myvspace{-10pt}
\end{figure}

\mysubsubsection{Discussions}
Readers might suspect that both intra- and inter-iteration sampling imbalance can be mitigated by modifying the sampling order. 
However, such approaches degrade model convergence because they incur randomness breaches when sampling the mini-batches during training. As shown in Figure~\ref{fig:loss}, we conduct a testbed to evaluate a length-based sampling method using Megatron-LM, which pre-sorts the dataset into three intervals (<4K, 4K–16K, and 16K–32K) and trains on them one by one. 
(By doing so, the sampling imbalance is mitigated, and the intervals can be trained with different parallel strategies since they differ in the maximum sequence lengths.) The results show that such a method suffers from a worse convergence compared to random sampling, and eventually, we observe a 2$\times$ gap in the perplexity metric. 
In contrast, Hydraulis preserves random sampling and synchronizes all gradients before model update, therefore the perplexity shows no significant deviation. This verifies the necessity of maintaining randomness in large model training.

Additionally, a prior work, HotSPa~\cite{ge2023hotspa}, addresses the data sampling imbalance in a different way---it performs length-based training within each iteration instead of over the entire dataset, in order to preserve the model convergence. 
By dynamically switching between expert-tuned, homogeneous parallel strategies for different sequence lengths, HotSPa accumulates gradients and performs a single global update per iteration. However, it lacks support for heterogeneous strategies and finer-grained data assignment, and frequent strategy switching introduces additional overhead.
We will empirically compare HotSPa and our work in \S\ref{sec:exp}.

\subsection{Data Packing Imbalance}

\begin{figure}[!t]
    \centering
    \includegraphics[width=\linewidth]{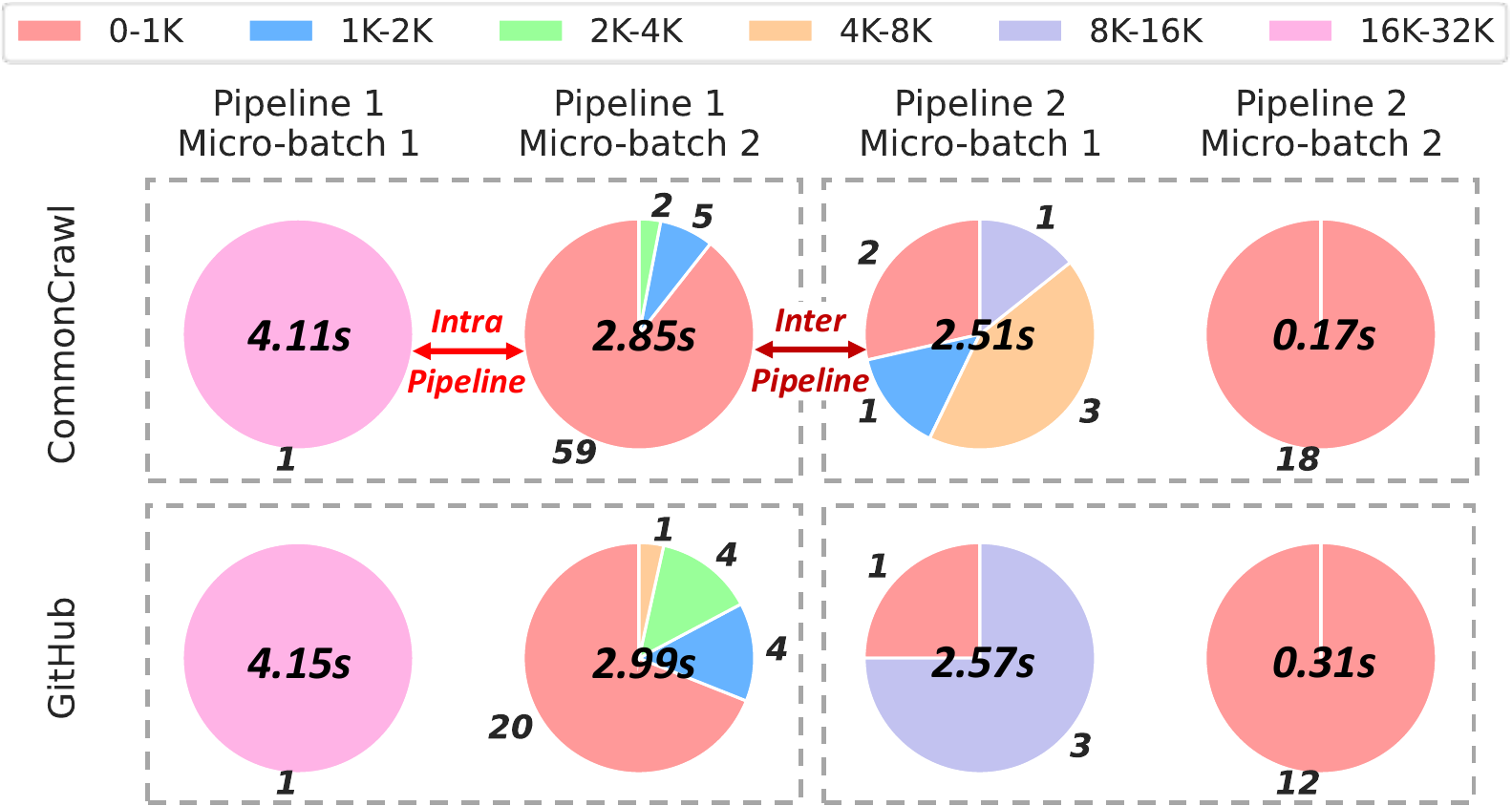}
    \caption{\small{The numbers of original sequences (classified by their lengths) in different micro-batches after packing (each micro-batch accounts for one packed sequence) and their corresponding  time cost. The experiment is conducted using a 13B LLaMA model with 16 Nvidia A800 GPUs.
    }}
    \label{fig:packing_imbalance}
    \myvspace{-10pt}
\end{figure}

We then focus on the imbalance issues of data packing. As illustrated in Figure~\ref{fig:packing_imbalance}, we observe significant distribution differences in the original sequences within packed sequences (i.e., micro-batches) due to varying sequence lengths. These variations occur both within micro-batches of the same pipeline and across different pipelines, resulting in execution time variance. Below, we analyze each aspect's impact separately.

\mysubsubsection{Intra-pipeline imbalance}
Since the memory consumption of each micro-batch is linear to the total length after packing, existing works commonly adopt max-length packing methods, like bin packing~\cite{yao1980binpacking, nvidia2023packing}, to balance memory usage. 
However, due to the quadratic time complexity of attention~\cite{vaswani2017attention, dao2022flashattention, nvidia2023packing}, the computation latency can hardly be balanced in this way--- 
for example with Figure~\ref{fig:padding_and_packing} again, attention latency can be nearly $20\times$ smaller when packing 32 sequences of 1K compared to a single 32K sequence.

In pipeline parallel (PP), this leads to uneven execution times across micro-batches, yet common pipeline scheduling, such as 1F1B~\cite{harlap2018pipedream} or zero-bubble~\cite{qi2023zerobubble}, requires uniform execution time of micro-batches to minimize pipeline bubbles. 
This highlights the importance of taking account of both the linear memory complexity and quadratic time complexity, rather than considering solely the memory as in max-length packing. To address this, we devise a nuanced packing strategy that considers the variation in original sequence lengths, striking a balance in terms of the execution time across different micro-batches within the memory constraint (\S\ref{subsec:sequence_packing}).

\mysubsubsection{Inter-pipeline imbalance}
Although different pipelines execute concurrently, the training latency is dictated by the slowest pipeline due to the need for gradient synchronization. Simply packing sequences to a fixed maximum length has a similar issue of unbalanced computation load. In some cases, if the sequence number after packing cannot be evenly distributed across pipelines, a more severe imbalance will arise. 
In contrast to current systems that first pack sequences and then attempt to distribute them evenly, we suggest reversing this process. Specifically, our work first dispatches sequences of varying lengths to different pipelines, and then focuses on packing them within each pipeline. This innovative dispatching paradigm ensures a more balanced load across pipelines, before organizing sequences into micro-batches (\S\ref{subsec:sequence_dispatching}).

%% file: sections/method.tex
% \myvspace{-8pt}
\section{Overview}
\label{sec:overview}

\begin{figure*}[!t]
\centering
\includegraphics[width=0.8\textwidth]{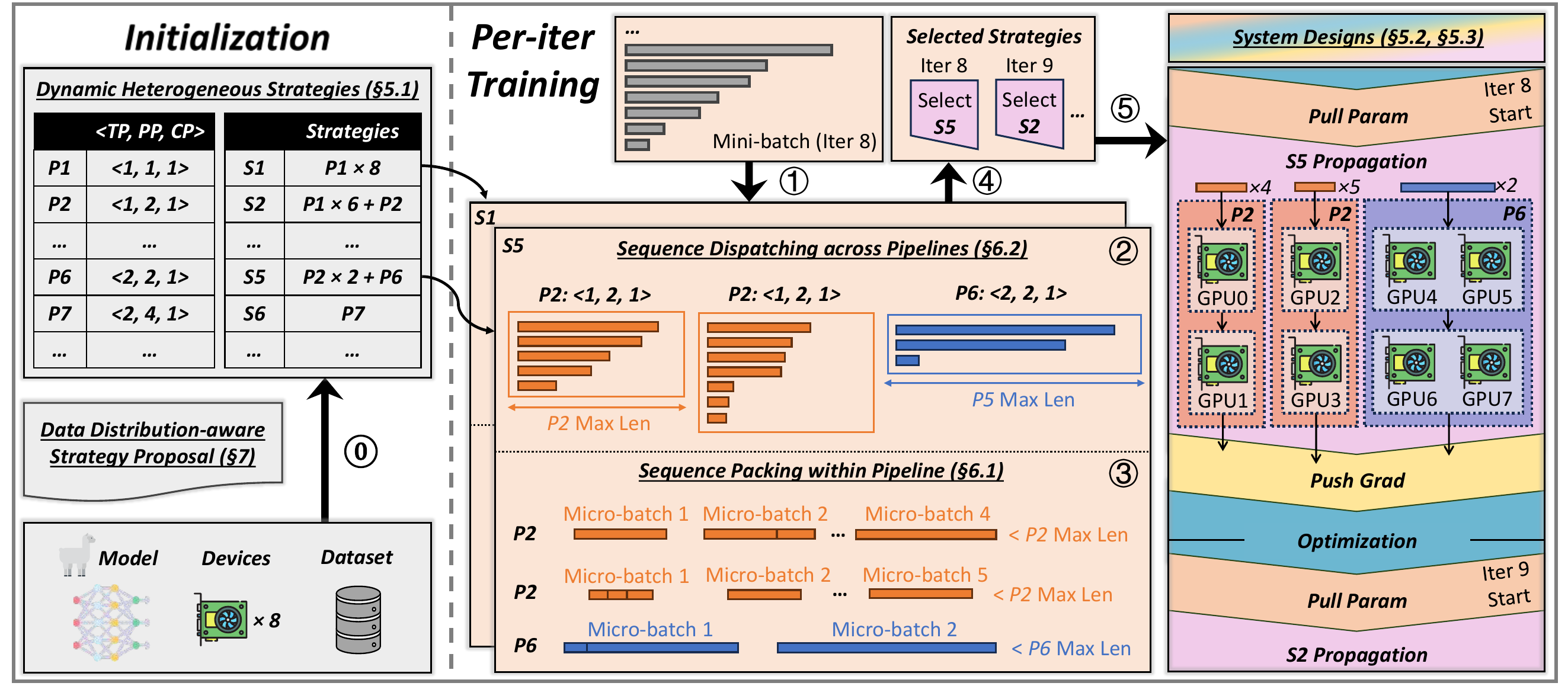}
\caption{\small{\system overview. The \textbf{gray} modules represent the initialization part, which is executed only once when training a specific model on fixed devices with a fixed dataset. The remaining modules operate per iteration during training: the \textbf{beige} modules indicate the two-stage sequence assignment, while the rest of the modules illustrate our system that supports dynamic heterogeneous strategies, with different colors representing different training phases. 
\system executes step {\textcircled{\scriptsize 0}} at initialization and steps {\textcircled{\scriptsize 1}}-{\textcircled{\scriptsize 5}} for each iteration.}}
\label{fig:overview}
\myvspace{-10pt}
\end{figure*}

We develop \system to address the training imbalances described above. Figure~\ref{fig:overview} illustrates the overview. 

To tackle the data sampling imbalance, we develop a brand new system that supports training with dynamic heterogeneous parallel strategies (\S\ref{sec:system_designs}). 
In each heterogeneous parallel strategy, the training pipelines can be associated with non-unique parallel schemes to differentiate between long and short sequences for each iteration, addressing the \textit{\ul{intra-iteration imbalance}}. 
And we further support adjusting the strategies dynamically across different iterations to tackle the \textit{\ul{inter-iteration imbalance}}.

To tackle the data packing imbalance, given a mini-batch of sequences and a strategy, we devise a two-stage assignment (\S\ref{sec:sequence_management}). 
The first stage determines how to dispatch sequences of varying lengths to different pipelines to solve the \textit{\ul{inter-pipeline imbalance}}. Then, the second stage meticulously packs the sequences within each pipeline to form different micro-batches, addressing the \textit{\ul{intra-pipeline imbalance}}. 
This two-stage assignment process also estimates the latency of the given strategy to train the given mini-batch.

The overall routine is as follows. 
At initialization (step {\textcircled{\scriptsize 0}}), we propose candidate strategies given the training task (\S\ref{sec:strategies_generator}). 
For each iteration of the training process, a mini-batch of sequences is drawn, based on which, we enumerate each candidate strategy (step {\textcircled{\scriptsize 1}}), employ the two-stage assignment to obtain its estimated performance (steps {\textcircled{\scriptsize 2}}-{\textcircled{\scriptsize 3}}), and select the best one (step {\textcircled{\scriptsize 4}}) for subsequent training (step {\textcircled{\scriptsize 5}}). 
The strategy selection (steps {\textcircled{\scriptsize 1}}-{\textcircled{\scriptsize 4}}) is done on CPUs, which overlaps with the training (step {\textcircled{\scriptsize 5}}) on GPUs.

\section{System Designs}
\label{sec:system_designs}

In this section, we start by mathematically defining our heterogeneous parallel strategies, then introduce the system techniques proposed to support these strategies and the dynamic transitions between them.

\subsection{Dynamic Heterogeneous Strategy Design}
\label{subsec:hydraulis_strategies}

\mysubsubsection{Definitions}
First, we focus on an independent \textbf{parallel scheme}, denoted as $P$, which describes the parallel configuration of one training pipeline. It can integrate multiple parallel methods. For example, if we employ tensor parallelism (TP) with a degree of 2 to partition individual model layers across devices, further combine it with pipeline parallelism (PP) of degree 4 to distribute these layers into sequential stages, and apply context parallelism (CP) of degree 1 to split input sequences, then $P$ can be represented as $\langle$TP,PP,CP$\rangle=\langle$2,4,1$\rangle$. In the following sections, to simplify representation, we introduce $\texttt{TP}(P)$, $\texttt{PP}(P)$ and $\texttt{CP}(P)$ to represent the TP, PP, and CP degrees for a given $P$, respectively. 
Based on this, we define the \textbf{parallel scheme space} $\mathbb{P}$, which contains various parallel schemes. For example with TP, PP, and CP again, $\mathbb{P}$ can be defined as 
\begin{equation*}
\begin{aligned}
\mathbb{P} = \big\{ P | 
&\texttt{TP}(P) \in \mathbb{D}_T, \texttt{PP}(P) \in \mathbb{D}_P, \texttt{CP}(P) \in \mathbb{D}_C, 
\\
&\texttt{TP}(P) \times \texttt{PP}(P) \times \texttt{CP}(P) \le N \big\},
\end{aligned}
\end{equation*}
where $\mathbb{D}_T = \mathbb{D}_C = \{1, 2, \cdots, N\}$, $ \mathbb{D}_P = \left\{ v \in \mathbb{N}^+ | v \text{ divides } \overline{\textup{layers}} \right\}$, $N$ is the number of GPUs, and $\overline{\textup{layers}}$ is the number of layers.

Next, we define our \textbf{heterogeneous strategy space} as the linear span of the parallel scheme space: $\mathbb{S} = \text{span}(\mathbb{P})$, meaning that each \textbf{heterogeneous strategy} can be represented as a linear combination of parallel schemes: $S = \sum_{i=1}^K d_i \times P_i$, which indicates that $S$ consists of $\sum_{i=1}^K d_i$ pipelines of $K$ different parallel schemes, with $d_i$ pipelines deployed with $P_i$. Figure~\ref{fig:overview} shows some example parallel schemes ($P$) and heterogeneous strategies ($S$) when the number of GPUs ($N$) is 8.

\mysubsubsection{Strategy execution and transition} During each training iteration, given a mini-batch of sequences, we select the most suitable $S$ to execute.\footnote{The selection in each iteration is based on the two-stage sequence assignment (\S\ref{sec:sequence_management}), and we propose the candidate strategies by analyzing the sequence length distribution of the entire dataset (\S\ref{sec:strategies_generator}).}
The pipelines will be assigned different sequences and execute forward and backward passes concurrently to compute gradients from the corresponding parameters. This process is termed the \textbf{propagation phase} (Figure~\ref{subfig:standard_op},~\ref{subfig:disaggregate_op}, step 3). Then, similar to traditional data parallelism (DP), all pipelines synchronize the model gradients and perform updates at the end of each iteration to obtain the new parameters and optimizer states. This process is called the \textbf{optimization phase} (Figure~\ref{subfig:standard_op},~\ref{subfig:disaggregate_op}, step 6).
Since different iterations inevitably select non-unique strategies, strategy transitions between iterations are necessary (e.g., in Figure~\ref{fig:overview}, iteration 8 uses strategy $S5$, while iteration 9 transitions to strategy $S2$).

\mysubsubsection{Challenges} To support the execution and transition of arbitrary heterogeneous strategies, the conventional approach to handling model
states must be revised. 
As shown in Figure~\ref{subfig:standard_op}, for current systems that employ a static, homogeneous parallel strategy, it is common practice to shard the model parameters and optimizer states according to the employed homogeneous strategy, and leverage all-gather and reduce-scatter operations to retrieve the parameters and synchronize the gradients, respectively. 
However, this approach falls short in  supporting dynamic heterogeneous strategies, as model partitioning varies significantly across pipelines. For example, in Figure~\ref{subfig:disaggregate_op}, pipelines with different TP degrees produce parameters and gradients with divergent partitioning granularities (i.e., tensors are split into non-uniform slices across GPUs). Consequently, it is infeasible to discover a single communication pattern to accomplish the parameter retrieval or gradient synchronization.
This limitation drives our development of the LNK (link) operator, which supports parameter retrieval and gradient
synchronization in an optimization-propagation disaggregated manner (\S\ref{subsec:op_disaggregation}), and we further
introduce subgraphs to encapsulate diverse communication patterns (\S\ref{subsec:subgraphs}). Furthermore, since gradient synchronization requires all pipelines to participate and is constrained by the slowest one, pipeline heterogeneity also poses significant load-balancing challenges, which we will address in \S\ref{sec:sequence_management}.

\subsection{Optimization-Propagation Disaggregation}
\label{subsec:op_disaggregation}

As illustrated in Figure~\ref{subfig:disaggregate_op}, we separate the optimization phase from the propagation phase, decoupling the GPU's role as a distributed storage from its computational role during propagation. 
Specifically, we keep the partitioning of FP32 parameters and optimizer states fixed for the optimization phase\footnote{We use a fully sharded strategy in the optimization phase, yet any other zero-redundancy sharding strategies are also feasible.}, while allowing an arbitrary parallel strategy to be used in the propagation phase. 
In such a disaggregated design, to ensure correct parameter partitioning during propagation as well as correct gradient partitioning for optimizer update, we establish two types of interaction:
(a) \textit{pull}, which retrieves the parameters from the optimization phase to an arbitrary heterogeneous parallel strategy (Figure~\ref{subfig:disaggregate_op}, steps 1–2); 
(b) \textit{push}, which synchronizes and scatters gradients back to the optimization phase (Figure~\ref{subfig:disaggregate_op}, steps 4-5). 
Compared to the all-gather and reduce-scatter in the conventional design that requires identical, homogeneous parallel strategy in the two phases (Figure~\ref{subfig:standard_op}, steps 1-2 and 4-5), our \textit{pull} and \textit{push} abstractions enable the use of arbitrary heterogeneous parallel strategies in the propagation phase as well as the seamless transition between different strategies.

\begin{figure}[!t]
    \centering
    \includegraphics[width=\linewidth]{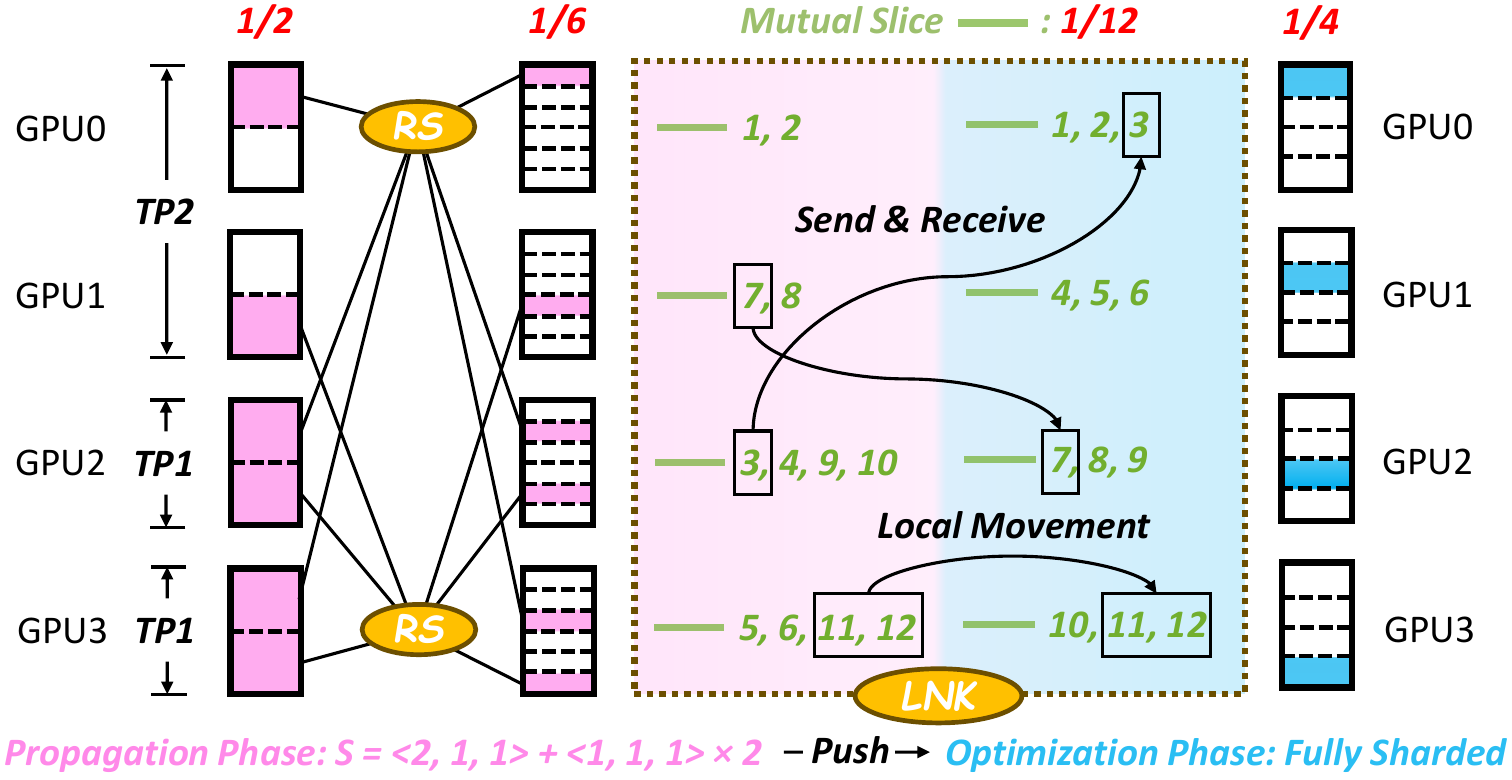}
    \caption{\small{Diagram of RS (reduce-scatter) and LNK (link) operations during \textit{push}. Pink and blue denote slices in the propagation and optimization phases, respectively, while green represents the \textit{mutual slice} (the finest-grained slice shared by both). The partitioning granularity is 4 in the propagation phase and 6 in the optimization phase (after reduce-scatter).} Thus, the \textit{mutual slice} has a granularity of 12 (i.e., the least common multiple of 4 and 6).}
    \label{fig:LNK}
    \myvspace{-10pt}
\end{figure}

To effectively manage \textit{pull} and \textit{push} interactions, we develop a specialized LNK (link) operator to handle complex communication patterns. As depicted on the right side of Figure~\ref{fig:LNK}, this operator utilizes three core primitives: \textit{send}, \textit{receive}, and \textit{local movement}, all operating on a fundamental unit called the \textit{mutual slice}, which represents the finest-grained data partition. These primitives enable dynamic resharding of parameters and gradients between propagation and optimization phases. For instance, in Figure~\ref{fig:LNK}, GPU1 generates a gradient slice during propagation, which is further partitioned into two finest-grained \textit{mutual slices} (7 and 8).
Since GPU0 requires \textit{mutual slice} 7 during the optimization
phase for model updates, it triggers a \textit{send} to
transfer \textit{mutual slice} 7 from GPU1 to GPU0. Because optimization states are uniquely partitioned across GPUs (without redundancy), the use of these primitives is deterministic.

Mathematically, assume there are $N$ GPUs, the optimization phase employs a fully sharded strategy, and the propagation phase uses a strategy $S = \sum_{i=1}^K d_i \times P_i$.
Let $\overline{TP}_\text{max} = \max_{i \in [1, K]} \texttt{TP}(P_i)$. For the \textit{pull} interaction, the \textit{mutual slices} are created by evenly partitioning the global parameter into  $\mathtt{LCM}\left(N, \overline{TP}_\text{max}\right)$ segments (where $\mathtt{LCM}$ denotes the least common multiple). Each GPU identifies its necessary \textit{mutual slices}. If a slice is local, it uses the \textit{local movement} operation; otherwise, it employs \textit{send} and \textit{receive} to exchange slices. Similarly, for the \textit{push} interaction, as shown on the left side of Figure~\ref{fig:LNK}, a reduce-scatter is first performed at the corresponding location to synchronize the gradient. Then the partitioning granularity of the \textit{mutual slice} is $\mathtt{LCM}\left(N, D \times \overline{TP}_\text{max}\right)$, where the additional factor $D = \sum_{i=1}^K d_i \times \texttt{CP}(P_i)$ accounts for gradients reduce-scatter across all model replicas. The LNK can be established still using the \textit{send}, \textit{receive} and \textit{local movement} primitives. Notably, the LNK mechanism operates independently of $\texttt{PP}(P_i)$, as PP doesn't partition individual parameters into slices. From the perspective of heterogeneous strategy execution, the LNK operator resides in different stages across training pipelines. Consequently, communication occurs only after the slowest pipeline reaches the LNK operator (the balancing across pipelines will be addressed in \S\ref{sec:sequence_management}).

It is noteworthy that our optimization-propagation disaggregation is fundamentally a generalization of current model states sharding techniques (i.e., ZeRO~\cite{rajbhandari2020zero} and FSDP~\cite{zhao2023fsdp})--- when using homogeneous strategies, the LNK operator in the \textit{pull} will be translated into an all-gather operation, and the LNK operator in the \textit{push} will be pruned (leaving a reduce-scatter operation), which are the same as current techniques.
For heterogeneous strategies, introducing the LNK operator incurs additional overhead to resolve the discrepancy between strategies. 
However, the \textit{send} and \textit{receive} communication can be overlapped with computation during propagation, and the \textit{local movement} incurs only minimal latency, so the overhead is small (as evaluated in \S\ref{subsec:performance_attribution}).

Last but not least, theoretically speaking, the communication during \textit{pull} and \textit{push} operations is bounded by the model size and does not scale with the number of GPUs. And the worst-case communication volume is exactly the same as that of current model states sharding techniques.
Furthermore, computation-communication overlap can be achieved when each micro-batch in the pipeline with $P_i$ contains more than $\texttt{TP}(P_i) \times \texttt{CP}(P_i) \times \frac{F}{B}$ tokens, where $F$ is the GPU’s half-precision compute capability (FLOPS), and $B$ is the slowest interconnect bandwidth (bytes/sec). 
This condition is independent of the model size or the number of GPUs. 
Due to the space constraint, detailed proof and analysis are provided in \ifappendix{Appendix~\ref{appendix:comm}}\else{Appendix B}\fi~\cite{hydraulis_appendix}.

\subsection{Subgraphs}
\label{subsec:subgraphs}

Although optimization-propagation disaggregation offers the ability to resolve the communication patterns needed for each heterogeneous strategy, it still poses the challenge of how to manage such diverse and complex optimization-propagation interactions, given that there would be numerous candidate heterogeneous strategies.
To tackle this challenge, we introduce subgraphs, which are subsets of the complete computation graph. 
The key observation is that different strategies only have unique patterns in some specific operators, while most user-defined computational operators remain unchanged. Therefore, subgraphs are developed to encapsulate these unique operators from different strategies for efficient representation, management and scheduling.   

\begin{figure}[!t]
    \centering
    \includegraphics[width=0.5\linewidth]{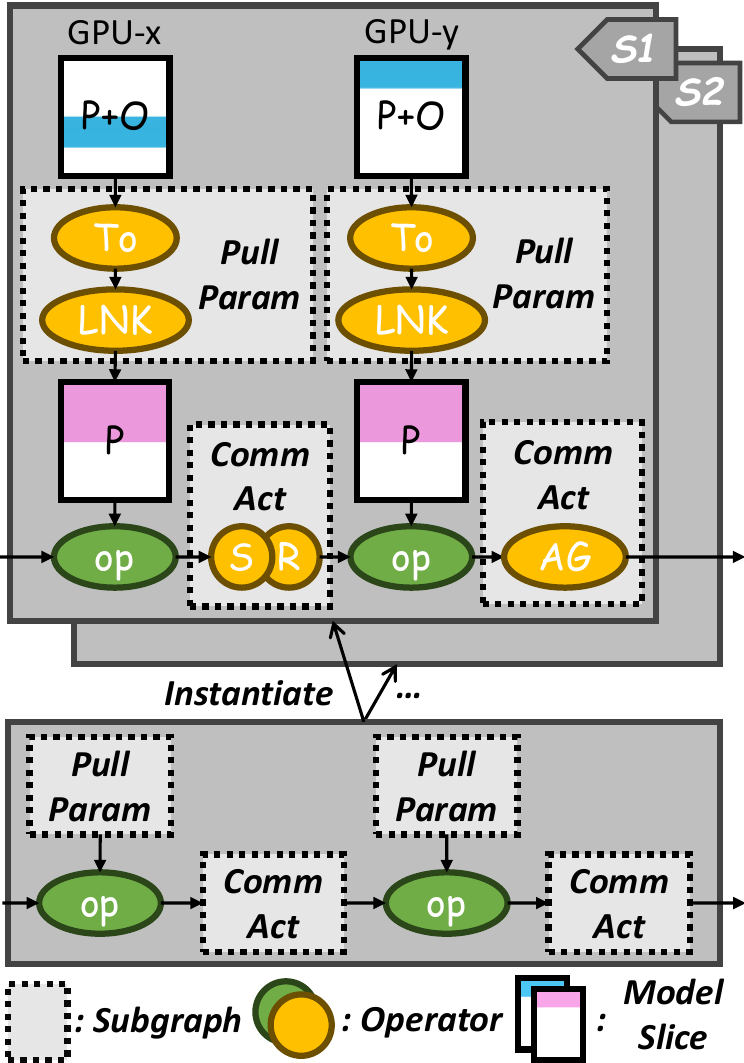}
    \caption{\small{The instantiation of \textit{Pull Parameter} and \textit{Communicate Activation} subgraphs during forward propagation. This involves inserting custom operators (e.g., send-receive, all-gather, LNK) to support arbitrary propagation strategies and optimization-propagation interactions.}}
    \myvspace{-10pt}
    \label{fig:subgraphs}
\end{figure}

In particular, we define three subgraph types: \textit{Pull Parameter}, \textit{Push Gradient}, and \textit{Communicate Activation}. Operators for interactions between optimization and propagation phases, like LNK, are placed in the first two, while activation-related communication operators (e.g., all-gather and reduce-scatter for TP, send-receive for PP) are included in the third.

\mysubsubsection{Subgraphs instantiation}
As illustrated in  Figure~\ref{fig:subgraphs}, users start by defining various computational operators to describe the training model. We then identify all potential points in the computational graph where different heterogeneous strategies might intervene and preserve several empty subgraphs (e.g., between transformer modules for potential pipeline parallel, before and after matrix multiplication for potential tensor parallel). Given a strategy
$S$, the user-defined operators are directly integrated into a new, strategy-specific graph. We then determine how to instantiate all predefined subgraphs with appropriate operators. For example, as depicted at the top of Figure~\ref{fig:subgraphs}, a data transfer operator and a strategy-specific LNK operator are added to instantiate a \textit{Pull Parameter} subgraph.
This instantiation process ultimately forms a complete graph for a specific $S$, enabling different GPUs to extract and execute their respective operators.

\begin{figure}[!t]
    \centering
    \includegraphics[width=\linewidth]{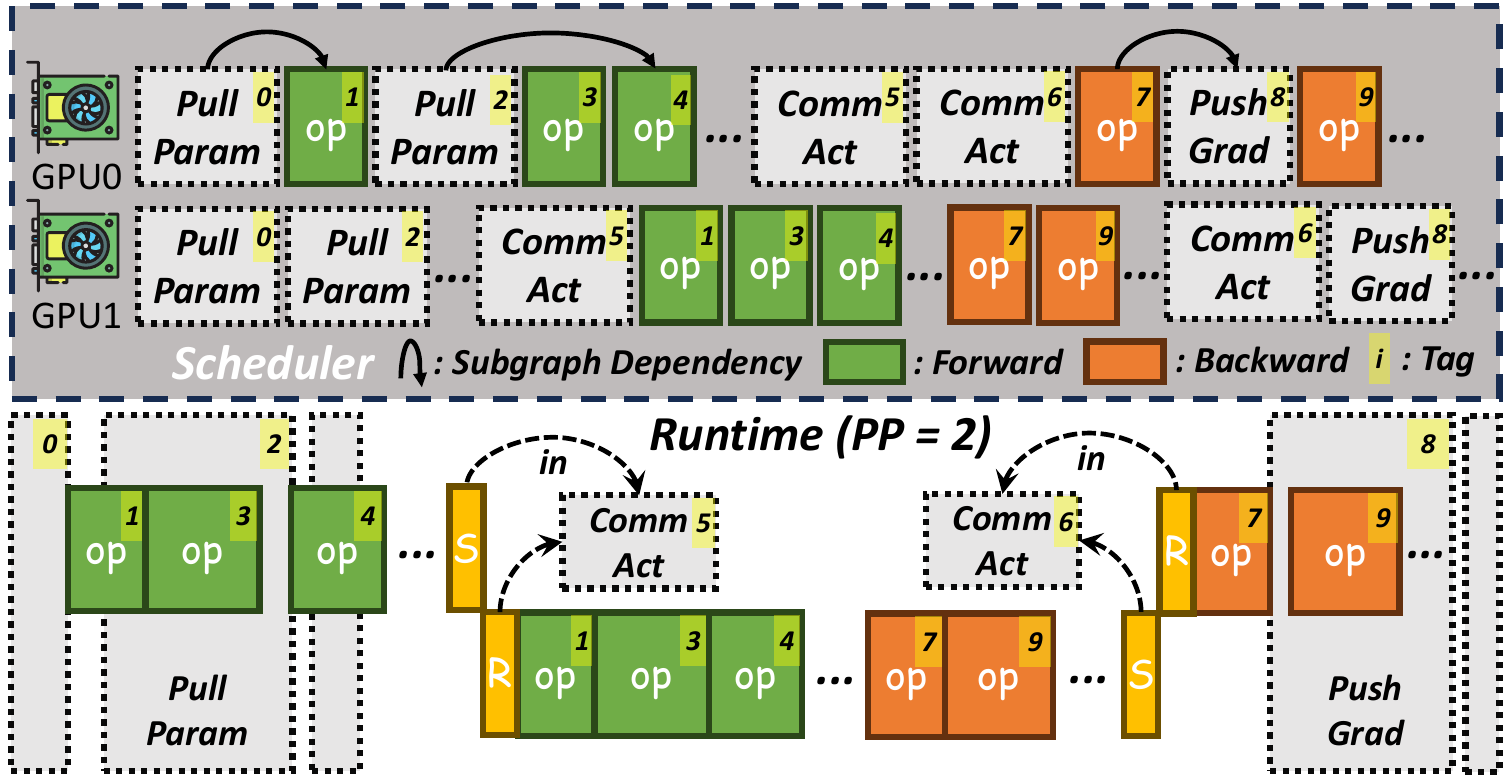}
    \caption{\small{Scheduling subgraphs with pipeline parallel (PP). In the first PP stage, \textit{Pull Parameter} subgraphs are scheduled immediately after the last operators requiring the parameter, and \textit{Push Gradient} subgraphs follow right after the operators that produce or accumulate the gradient. In subsequent PP stages, \textit{Pull Parameter} and \textit{Push Gradient} are grouped at the beginning and end, respectively, to utilize inherent PP bubbles. This scheduling scheme effectively hides communication latency within the \textit{Pull Parameter} and \textit{Push Gradient} subgraphs.}}
    \label{fig:scheduler}
    \myvspace{-10pt}
\end{figure}

\mysubsubsection{Subgraphs scheduling}
During execution, we optimize overlap by controlling the execution order of subgraphs and computational operators.
Each subgraph can be treated as a cohesive unit and integrated into the scheduler as a fused operator.
As illustrated in Figure~\ref{fig:scheduler}, by using a typical \texttt{cudaStream} to execute the subgraphs and interleaving them with computational operators, we enable communication within subgraphs to overlap with computation, thereby reducing latency.

\section{Two-stage Sequence Assignment}
\label{sec:sequence_management}

In this section, we present our two-stage sequence assignment approach. Specifically, given a mini-batch of sequences and an arbitrary strategy, we establish an optimization problem to pack sequences precisely, and ensure balanced sequence dispatching based on load estimates of each pipeline, addressing both intra- and inter-pipeline imbalances. 
In each iteration, we apply the assignment to all candidate strategies, estimate the propagation latency, and select the optimal one.

\subsection{Sequence Packing within Pipeline}
\label{subsec:sequence_packing}

We first introduce the packing strategy within each pipeline. 
Assume there are $U$ sequences with lengths $\{l_i\}_{i=1}^{U}$ assigned to a pipeline configured with parallel scheme $P$. 
To pack these sequences into multiple micro-batches, there are two essential terms to determine: the number of micro-batches to craft (denoted as $V$) and how each sequence should be assigned.

\mysubsubsection{Cost Model} Following prior works on automatic parallel training~\cite{zheng2022alpa, lihaoyang2025malleus,lobra,memo}, we develop two profiling-based cost models to estimate memory usage and latency. For a given $P$, there is a maximum sequence length it can support, denoted as $\texttt{MaxLen}(P)$ (i.e., sequence exceeding this length will cause OOM on $P$), which can be determined in advance using our memory model (detailed in 
\ifappendix{Appendix~\ref{appendix:memory_model}}\else{Appendix C.1}\fi~\cite{hydraulis_appendix}). 
Additionally, we introduce a latency model, $\texttt{T}(l, P)$, which estimates the time taken to run a sequence of length $l$ on $P$ for a single forward and backward propagation of a single pipeline stage (detailed in \ifappendix{Appendix~\ref{appendix:latency_model}}\else{Appendix C.2}\fi~\cite{hydraulis_appendix}).
Similar to prior works, re-profiling is required for different model architectures but is performed as a one-shot process per architecture. This overhead is worthwhile given the significant speedup in large-scale training. Moreover, this process is independent of the cluster size, as scaling
the cluster only increases the number of pipelines, without affecting the set of valid $P$ or the sequence length $l$ used during training.

\mysubsubsection{Solving for one $V$}
The problem is easy to solve if the number of micro-batches $V$ is given. 
Specifically, as mentioned in \S\ref{subsec:data-sampling}, by employing the varlen functionality of FlashAttention v2~\cite{dao2022flashattention,dao2023flashattention2}, the latency of a packed sequence can be approximated as the sum of latency for each original sequence. 
Meanwhile, to address the intra-pipeline imbalance, our goal is to minimize the duration of the longest-running micro-batch. 
Thus, we can formulate the following problem on top of the minimum sum partition (MSP) problem:
\begin{equation}
\small
\begin{aligned}
\argmin_{o_{i,j}}
& \max\limits_{1 \le j \le V}{\left(\sum\limits_{1 \le i \le U}{o_{i,j} \times \texttt{T}\left(l_i, P\right)}\right) \times (\texttt{PP}(P) - 1 + V)} \\
\text{s.t.} & \sum\limits_{1 \le j \le V}{o_{i,j}} = 1, o_{i,j} \in \left\{0, 1\right\}, \forall i \in \left[1, U\right] \\
&\sum\limits_{1 \le i \le U}{o_{i,j} \times l_i} \le 
\texttt{MaxLen}\left(P\right)
, \forall j \in \left[1, V\right] 
\end{aligned}
\label{eq:packing_ilp}
\end{equation}
where $o_{i,j} \in \{0,1\}$ indicates whether the $i$-th sequence is packed into the $j$-th micro-batch. 
Compared to the MSP problem, the problem defined in Eq.~\eqref{eq:packing_ilp} additionally ensures that the length after packing must not exceed $\texttt{MaxLen}(P)$, and the term $\texttt{PP}(P) - 1 + V$ represents the propagation latency of the entire pipeline.

\mysubsubsection{Enumeration of $V$}
To locate the optimal number of micro-batches $V^*$, we can simply enumerate the value for $V \in [1,U]$, solve Eq.~\eqref{eq:packing_ilp}, and choose the best one. 

Note that Eq.~\eqref{eq:packing_ilp} is an Integer Linear Programming (ILP) problem, but its complexity is only related to the number of sequences assigned to $P$ (i.e., $U$) and the number of micro-batches (i.e., $V$). These do not increase with cluster size, and the enumeration can be done in parallel, ensuring fast problem-solving even on large-scale clusters (detailed in \ifappendix{Appendix~\ref{appendix:planning_time}}\else{Appendix F}\fi~\cite{hydraulis_appendix}). 

\subsection{Sequence Dispatching across Pipelines}\label{subsec:sequence_dispatching}

Next, we consider how the sequences in a mini-batch should be dispatched across the pipelines to address the inter-pipeline imbalance. A na\"ive approach is to bucket sequences by length. However, given the long-tail distribution of sequence lengths (where long sequences are rare, as shown in Figure~\ref{fig:distribution}), pipelines configured for long sequences may receive too few assignments during iterations dominated by short sequences. To improve load balancing, we need a more flexible approach that allows short sequences to be processed by any pipeline.
Consider a specific strategy $S = P_1 + \ldots + P_D$ satisfying $\texttt{MaxLen}\left(P_1\right) \ge \ldots \ge \texttt{MaxLen}\left(P_D\right)$. Assume a mini-batch contains $B$ 
sequences with lengths $\{l_i\}_{i=1}^B$. 
We denote $J_i$ as the number of pipelines that the $i$-th sequence can run on, 
satisfying $\texttt{MaxLen}\left(P_{J_i}\right) \ge l_i > \texttt{MaxLen}\left(P_{J_i + 1}\right)$. Thus, $l_i$ can be dispatched to any of $P_1, \ldots, P_{J_i}$. Since the pipelines execute concurrently, our goal is to dispatch the $B$ sequences across the $D$ pipelines while minimizing the propagation latency of the longest-running pipeline. 
 
However, the propagation latency of each pipeline cannot be directly computed without solving the packing within it (i.e., \S\ref{subsec:sequence_packing}). 
Fortunately, it is feasible to estimate the lower bound of latency. 
In particular, let 
$m_{i,j} \in \{0,1\}$
indicate whether the $i$-th sequence is assigned to $P_j$, 
then the lower bound of running time of the $j$-th pipeline can be expressed as (detailed derivation in \ifappendix{Appendix~\ref{appendix:derivation_of_lower_bound}}\else{Appendix E.1}\fi~\cite{hydraulis_appendix}):
\begin{equation}
\small
\sum_{i=1}^B m_{i,j} \times \texttt{T}(l_i, P_j) + \texttt{T}\left(\max_{1 \le i \le B}\left\{m_{i,j} \times l_i\right\}, P_j\right) \times \left(\texttt{PP}\left(P_j\right) - 1\right) 
\label{eq:dispatching_lower_bound}
\end{equation}
Denote Eq.~\eqref{eq:dispatching_lower_bound} as $\texttt{LowerBound}(\{m_{i,j}\}_{i=1}^B, \{l_i\}_{i=1}^B, P_j)$, then the dispatching can be formulated as a (0,1)-ILP problem:
\begin{equation}
\small
\begin{aligned}
\argmin_{m_{i,j}}
& \max\limits_{1 \le j \le D}
\texttt{LowerBound}(\{m_{i,j}\}_{i=1}^B, \{l_i\}_{i=1}^B, P_j) \\
\text{s.t.} & \sum\limits_{1 \le j \le J_{i}}{m_{i,j}} = 1, \forall i \in \left[1, B\right] \\
& m_{i,j} \in \left\{0, 1\right\}, \forall i \in \left[1, B\right], \forall j \in \left[1, D\right]
\end{aligned}
\label{eq:dispatching_ilp}
\end{equation}
The number of decision variables of Eq.~\eqref{eq:dispatching_ilp} is linear w.r.t. number of sequences in a mini-batch (i.e., $B$) and the number of pipelines in the strategy (i.e., $D$). 
In practice, as the cluster scale expands, both will increase proportionally, making the assignment problem-solving cost unacceptable in large-scale scenarios. 
Fortunately, there are extensive studies about how to approximately solve assignment problems in polynomial time with a guaranteed error bound~\cite{duan2014linear, shmoys1993approximation, chen2020online}. 
Thus, we follow previous works to solve Eq.~\eqref{eq:dispatching_ilp} approximately, which involves running multiple random trials to greedily dispatch sequences and selecting the best trial. Empirical results demonstrate that the approximation errors are consistently lower than 10\% compared to directly solving Eq.~\eqref{eq:dispatching_ilp}.
More details about our approximate solution can be found in \ifappendix{Appendix~\ref{appendix:dispatching_datails} and Appendix~\ref{appendix:planning_time}}\else{Appendix E and Appendix F}\fi~\cite{hydraulis_appendix}

\section{Data Distribution-aware Strategy Proposal}
\label{sec:strategies_generator}

Ideally, for each iteration, we could explore all possible strategies, estimate their propagation latency via the two-stage sequence assignment, and find the optimal one.
However, given the vast size of the strategy space $\mathbb{S}$, this is impractical. Thus, we analyze the dataset in advance and propose a refined subset $\hat{\mathbb{S}} \subset \mathbb{S}$, which only contains the most promising candidates.

\mysubsubsection{Problem definition}

Our proposal is based on two considerations regarding intra- and inter-iteration imbalances: First, since intra-iteration imbalance typically reflects the expected distribution of the dataset, we aim to preserve strategies that demonstrate an overall advantage over the entire dataset. Second, due to inter-iteration imbalance, characterized by variations in maximum sequence lengths, we aim to identify non-unique strategies across different maximum lengths.
Then, we have the following problem.

\begin{problem}
Given the number of GPUs $N$, the dataset $\mathbb{D}$, and an arbitrary sequence length $L$. With sequence lengths randomly sampled within $[0, L]$ over $\mathbb{D}$, we seek the optimal heterogeneous strategy $S^*_{L \vert \mathbb{D}}$ from $\mathbb{S}$ that minimizes the expected cost. Then the proposed subset is defined as the union: $\hat{\mathbb{S}} = \bigcup_{L \leq L_{\text{max}}} \{ S^*_{L \vert \mathbb{D}} \}$, 
where $L_{\text{max}}$ is the maximum sequence length to support (i.e., context length).
\label{problem:1}
\end{problem}

\noindent We first simplify the problem by forcing each pipeline to process only the sequences within a specific length interval, rather than any arbitrary lengths. 
This allows the problem to be solved via dynamic programming. 
Later we lift the restriction by continuous relaxation, enhancing the solution. 

\mysubsubsection{Dynamic programming}
We define $t[n][l]$ as the minimum achievable propagation latency using up to $n$ GPUs to process all sequences in the dataset with lengths not exceeding $l$. Our goal is to determine the value $t[N][L]$.
The initial states are obvious that $t[n][0] = 0, \forall n > 0$ and $t[0][l]=\infty, \forall l > 0$. 
Next, we focus on the state transition for dynamic programming.
Let $\mathbb{D}_{(l-l^\prime, l]}$ denote the sub-dataset containing sequences within the interval $(l-l^\prime, l]$, and assume the parallel scheme space comprises $K$ unique schemes $\{P_k\}_{k=1}^K$. Based on our simplification, we only need to consider assigning this sub-dataset to one kind of scheme. 
Suppose it is assigned to $d$ pipelines of $P_k$ and $P_k$ requires $\texttt{N}\left(P_k\right)$ GPUs, then this assignment consumes $d \times \texttt{N}\left(P_k\right)$ GPUs in total. Besides, we must ensure $\texttt{MaxLen}\left(P_k\right)\ge l$. Then the propagation latency of this sub-dataset can be approximated as $\frac{1}{d}\sum_{x \in \mathbb{D}_{(l-l^\prime, l]}}\texttt{T}(x, P_k)$. 
Meanwhile, the minimum propagation latency for the remaining sequences is $t\left[n-d \times \texttt{N}\left(P_k\right)\right]\left[l-l^\prime\right]$. Therefore, we have the following state transition expression:
\begin{equation}
\small
\begin{aligned}
&t[n][l] = \min\left(t[n-1][l], \min_{(k,d) \in {Cond}_{n,l}, l^\prime < l} \left\{{T}_{k,d,l^\prime}\right\}\right), \ \textup{where} \\
&{T}_{k,d,l^\prime} = \max\left(t[n-d \times \texttt{N}\left(P_k\right)][l-l^\prime],\frac{1}{d}\sum_{x \in \mathbb{D}_{(l-l^\prime, l]}}\texttt{T}(x, P_k)\right) \\
&{Cond}_{n,l} = \left\{ (k, d) \mid \texttt{MaxLen}\left(P_k\right) \ge l, \, d \times \texttt{N}\left(P_k\right) \le n \right\} \\
\end{aligned}
\end{equation}
The use of $\max(\cdot,\cdot)$ in $T_{k,d,l^\prime}$ is because the slowest pipeline determines the overall latency. Based on the dynamic programming, we also obtain the corresponding optimal strategy $S[n][l]$ upon the calculation of $t[n][l]$, and $S[N][L]$ will be viewed as the solution $S^*_{L \vert \mathbb{D}}$ under $L$ in Problem 1.

By leveraging the optimal sub-structure property of dynamic programming, once we determine
$S[N][L_\text{max}]$, we obtain
$S[N][L]$ for all $L \le L_\text{max}$. We can then select all unique $S[N][L]$ from $L=1$ to $L=L_{max}$ to form $\hat{\mathbb{S}}$. 

\begin{figure}[!t]
    \centering
    \includegraphics[width=0.9\linewidth]{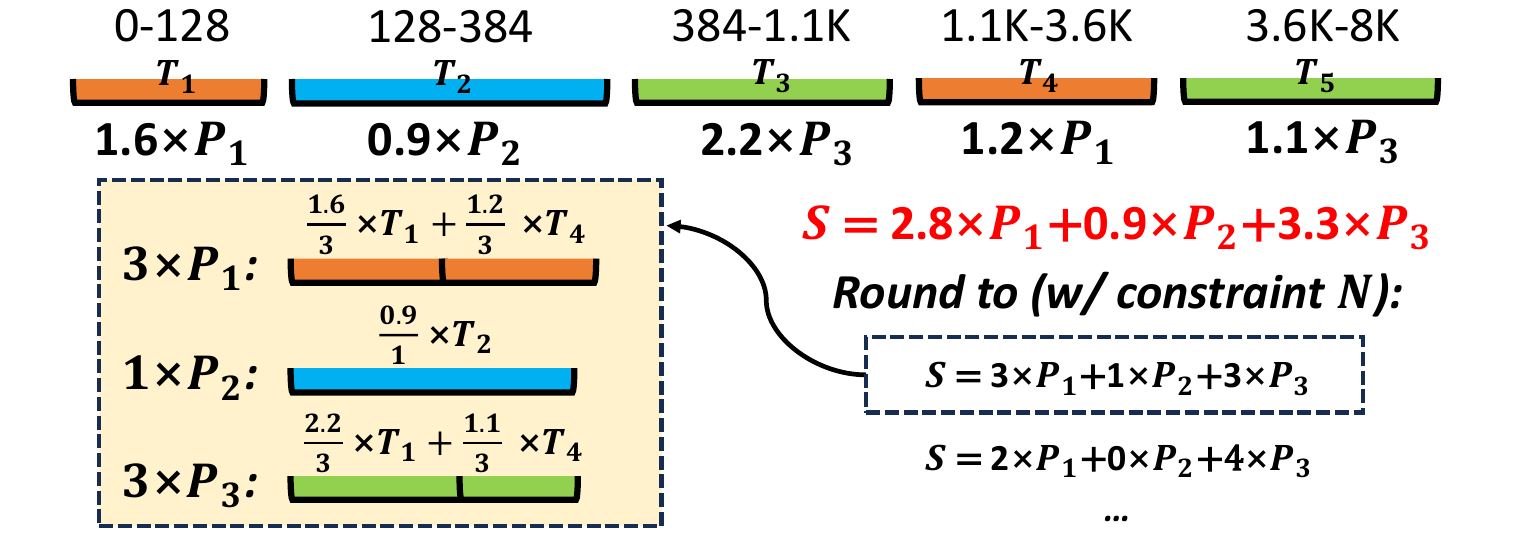}
    \caption{\small{Dynamic programming with non-integer values. Different colors represent various $P$. The segments indicate sequences within each dataset interval, with their lengths corresponding to latency. Rounding the number of $P$ requires scaling the results.}}
    \label{fig:generator}
    \myvspace{-10pt}
\end{figure}

\mysubsubsection{Continuous relaxation}
Based on the dynamic programming, we lift the restriction that each pipeline can only handle a sequence over a single length interval. Specifically, we execute the dynamic programming with non-integer values for $n$ and $d$. This means sequences in a length interval can use ``a portion'' of a pipeline, enabling a complete pipeline to support multiple intervals. Consequently, the final solution $S[N][L]$ might involve having non-integer numbers, as exemplified in Figure~\ref{fig:generator}. In such cases, we round and adjust to include all nearby integer solutions in $\hat{\mathbb{S}}$, while ensuring that the total number of GPUs does not exceed $N$.\footnote{The whole algorithm's worst-case time complexity is $O(KN^2L_\text{max}^2)$. In our experiments, we set the step size of $n$ and $d$ to 0.1 and $l$ to 128, with the actual time cost detailed in  \ifappendix{Appendix~\ref{appendix:planning_time}}\else{Appendix F}\fi~\cite{hydraulis_appendix}.}

Undoubtedly, even though the original non-integer solution achieves optimal results, rounding compromises the optimality. Additionally, the data distribution in different iterations does not perfectly align with that of the entire dataset. Due to these limitations, we can only suggest potential strategy candidates, whereas the two-stage sequence assignment is still necessary for selecting optimal strategies.

%% file: sections/exp.tex
\begin{figure*}[!t]
    \centering
    \includegraphics[width=1.0\textwidth]{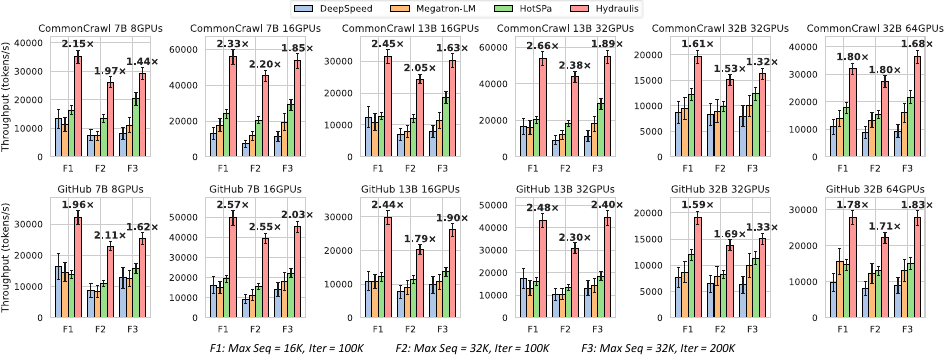}
    \caption{\small{End-to-end evaluation. We test the workloads of three configurations, F1-F3, in each scenario. ``Max Seq'' represents the number of tokens in the longest sequence, and ``Iter'' denotes the number of tokens sampled (a.k.a. batch size) in a single iteration. Error bars indicate the standard deviation across iterations.}}
    \label{fig:e2e}
    \myvspace{-10pt}
\end{figure*}

\section{Evaluation}
\label{sec:exp}

\begin{figure}[!t]
    \centering
    \includegraphics[width=\linewidth]{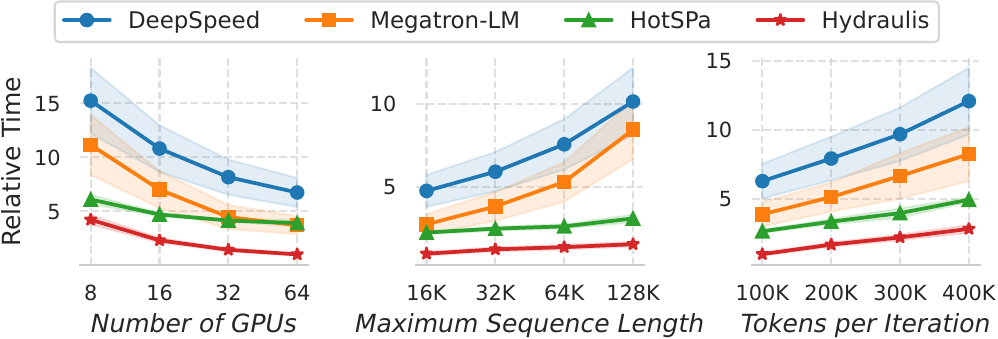}
    \caption{\small{Comparison of scalability across various factors, using the fastest one as the reference for relative changes. Shaded areas indicate the standard deviation across iterations.}}
    \label{fig:scalability}
    \myvspace{-10pt}
\end{figure}

\begin{figure*}[!t]
    \begin{subfigure}{0.71\linewidth}
        \centering
        \includegraphics[width=\linewidth]{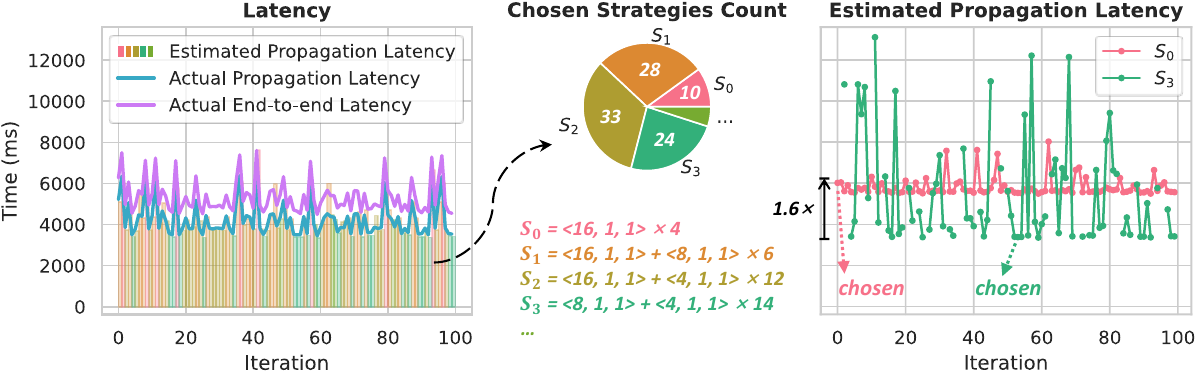}
        \subcaption{\small{Case study w.r.t parallel strategies.}}
        \label{subfig:case_study_1}
    \end{subfigure}
    \begin{subfigure}{0.285\linewidth}
        \centering
        \includegraphics[width=\linewidth]{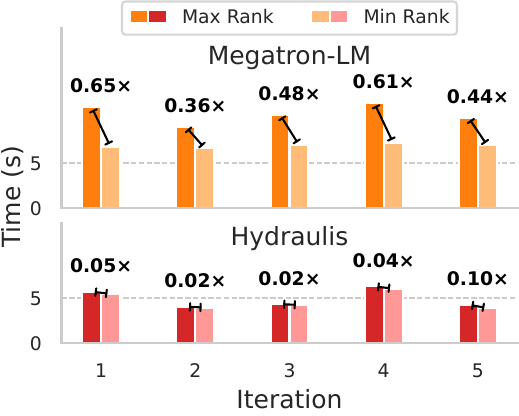}
        \subcaption{\small{Case study w.r.t sequence assignment.}}
        \label{subfig:case_study_2}
    \end{subfigure}
    % \myvspace{-15pt}
    \caption{\small{(a) Latency over 100 training iterations, with colors indicating the strategy selected for each iteration. (b) Comparison of load balancing between \system and Megatron-LM. We present the maximum and minimum propagation latency among all ranks for 5 iterations.}}
    \label{fig:case_and_ablation}
    \myvspace{-10pt}
\end{figure*}

\subsection{Experimental Setup}
\label{subsec:exp_setup}

\mysubsubsection{Environment}
Our experiments are conducted with 8 nodes, each containing 8 Nvidia A800 GPUs, with a total number of 64 GPUs. GPUs within a node are connected via NVLink with a bandwidth of 400 GB/s, and nodes are interconnected using Infini-Band with a bandwidth of 200 GB/s.

\mysubsubsection{Baselines}
We compare \system with Megatron-LM~\cite{shoeybi2019megatron, narayanan2021efficient, korthikanti2022reducing}, DeepSpeed~\cite{rasley2020deepspeed, jacobs2023deepspeedulysses}, and HotSPa~\cite{ge2023hotspa}. 
Both Megatron-LM and DeepSpeed employ a single, static homogeneous strategy throughout the training process. 
HotSPa uses heuristics to configure multiple homogeneous strategies and enables switching between them within every iteration. However, due to the overhead of switching, HotSPa is limited to setting 3-4 strategies. 
We enumerate all competitors' parallel strategies and select the optimal one to maximize baseline performance (see \ifappendix{Appendix~\ref{appendix:configuration}}\else{Appendix G}\fi~\cite{hydraulis_appendix} for details).

\mysubsubsection{Models and datasets}
Most existing large models are based on the decoder-only Transformer architecture. We select the widely used LLaMA2 series models and conduct experiments with 7B, 13B and 32B model sizes~\cite{touvron2023LLaMA}, using CommonCrawl~\cite{tiiuae2024falconrefinedweb} and GitHub~\cite{codeparrot2024githubcode}, which are widely used open-source datasets and have served as the major training corpora for many prestigious Transformer models. 

\mysubsubsection{Worloads and metrics}
By default, we sample 100K tokens per iteration\footnote{We consider 100K tokens per iteration for our 64-GPU cluster, scaled proportionally from the 16M used in LLaMA with an 8192-GPU cluster~\cite{dubey2024llama3}.} with a 32K context length, and use the Adam optimizer~\cite{kingma2015adam}. 
We conduct 20 warm-up iterations and then summarize the throughput (tokens per second) or per-iteration latency over 100 iterations. 

\subsection{End-to-end Experiments}
\label{subsec:e2e_exp}

We first evaluate \system against three baselines across various configurations (baselines' strategies are detailed in \ifappendix{Appendix~\ref{appendix:configuration}}\else{Appendix G}\fi~\cite{hydraulis_appendix}). The results in Figure~\ref{fig:e2e} demonstrate that \system achieves 1.32-2.66$\times$ improvement over the best baselines.

As DeepSpeed and Megatron-LM are limited to a single fixed strategy, they must adopt less efficient strategies to accommodate the memory consumption of long sequences, degrading the overall efficiency given most sequences are relatively short. 
Although HotSPa is capable of using multiple homogeneous strategies within an iteration to train sequences of varying lengths, it requires multiple times of model hot switching. 
This incurs apparent overheads, especially when per-iteration time is short (e.g., in the ``GitHub 7B 8GPUs F1'' scenario, the average iteration time is merely 6.7s, while nearly 2s is spent on hot switching).\footnote{A more detailed analysis of HotSPa's switching overhead can be found in \ifappendix{Appendix~\ref{appendix:hotspa}}\else{Appendix H}\fi~\cite{hydraulis_appendix}.}
On the contrary, \system handles variable-length sequences using dynamic heterogeneous strategies, while eliminating the overhead of strategy transition via disaggregating optimization and propagation.

Moreover, none of the baselines consider fine-grained sequence management. This problem is particularly severe for HotSPa--- in some cases, there are only 1-2 long sequences per iteration, yet HotSPa is forced to adopt an individual strategy with all GPUs to process such long sequences, leaving a few GPUs with no sequences to process
(e.g., in the ``CommonCrawl 32B 64GPUs F2'' scenario, HotSPa adopts $\left<TP,PP,CP\right> \times DP=\left<16,1,1\right>\times4$ to process sequences ranging 16K-32K in length, often resulting in 3/4 or 1/2 GPUs being completely unused). 
Instead, \system, with the two-stage sequence assignment, meticulously organizes the training workloads of different sequences for better performance, as we will evaluate more in-depth later in \S\ref{subsec:performance_attribution}.

\subsection{Scalability}
\label{subsec:scalability}

We then assess the scalability of different systems by examining three factors: (a) number of GPUs, (b) maximum sequence length, and (c) tokens per iteration. Using the LLaMA 7B model and the CommonCrawl dataset, we start with a default setup of 16 GPUs, 32K maximum sequence length, and 200K tokens per iteration. By systematically altering one factor while maintaining others, we gain insights into each factor's impact. Results are presented in Figure~\ref{fig:scalability}. Additionally, we analyze how (a) data imbalance and (b) problem-solving cost scale with increasing training workloads, particularly when both the cluster size (number of GPUs) and batch size (tokens per iteration) grow proportionally.

\mysubsubsection{Assessment: number of GPUs}
As the number of GPUs increases, all systems experience a sub-linear reduction in latency due to the communication overhead introduced by parallelism. Specifically, for HotSPa, the communication overhead from hot switching does not decrease with the number of GPUs, causing its overall performance to be surpassed by Megatron-LM when using 64 GPUs. In contrast, \system introduces minimal additional communication overhead (only the LNK operators), which can overlap with computation. Consequently, its latency consistently outperforms other baselines.

\mysubsubsection{Assessment: maximum sequence length}
As the maximum sequence length grows, DeepSpeed and Megatron-LM resort to less efficient strategies, resulting in a significant increase in latency. Notably, when Megatron-LM scales from 64K to 128K, it must enable cross-node TP, which greatly increases latency. In contrast, by customizing parallel strategies for sequences of different lengths, HotSPa and \system are less sensitive to maximum sequence length changes, with \system demonstrating superior performance over HotSPa.

\mysubsubsection{Assessment: tokens per iteration}
As the number of tokens per iteration increases, all systems experience an almost linear increase in latency. However, HotSPa and \system exhibit a slightly lower growth rate, as they take account of the long-tail distribution of sequence lengths within iterations. 

\mysubsubsection{Analysis: data imbalance}
When scaled to more GPUs, each iteration typically samples more training data, which may mitigate inter-iteration imbalance because longer sequences become more likely to be selected. 
To assess this, we conduct a 1024-GPU simulation with 16M tokens sampled per iteration (detailed in \ifappendix{Appendix~\ref{appendix:larger_batch_size}}\else{Appendix A.1}\fi~\cite{hydraulis_appendix}).
The results show that the maximum sequence length fluctuates between 16K-32K, indicating that inter-iteration imbalance still remains. 
Meanwhile, intra-iteration imbalance persists due to the dataset's inherent sequence length distribution, while intra- and inter-pipeline imbalances originate from the discrepancy between the attention mechanism’s quadratic time complexity and linear memory complexity relative to sequence length. Since these imbalances are scale-independent, they persist at larger scales.

\mysubsubsection{Analysis: problem-solving cost}
We also conduct simulation experiments to evaluate the scalability of problem solving when scaled to 1024 GPUs (detailed in \ifappendix{Appendix~\ref{appendix:planning_time}}\else{Appendix F}\fi~\cite{hydraulis_appendix}), which show that the problem-solving cost is well-bounded. 
To be specific, the strategy proposal (\S\ref{sec:strategies_generator}) incurs acceptable overhead (< 2 minutes) as it executes only once prior to training. Meanwhile, both sequence packing (\S\ref{subsec:sequence_packing}) and dispatching (\S\ref{subsec:sequence_dispatching}) operate on CPUs and can be fully overlapped with GPU training (i.e., the problem-solving for subsequent iterations occurs concurrently with current iteration training). 
Furthermore, larger cluster sizes provide additional CPU resources to increase the concurrency of problem solving, allowing us to further amortize the solving time per iteration.

\begin{figure}[!t]
    \centering
    \includegraphics[width=\linewidth]{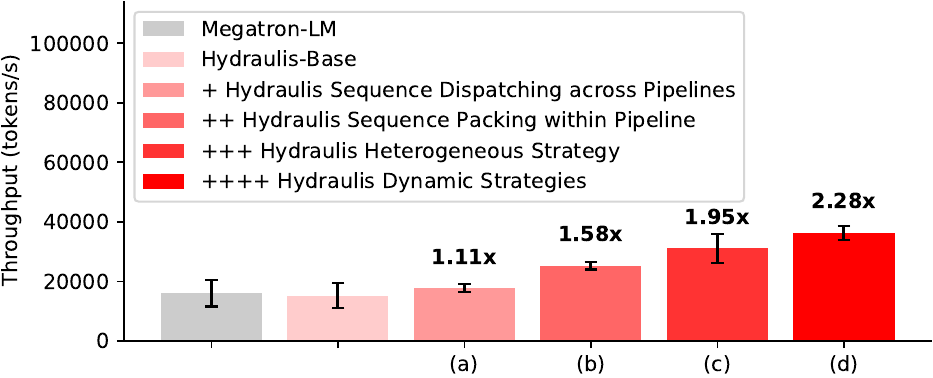}
    \caption{\small{Ablation study. Error bars indicate the standard deviation across iterations. Total throughput changes as different components of \system are incrementally integrated.}}
    \label{fig:ablation}
    \myvspace{-10pt}
\end{figure}

\subsection{Performance Interpretation}
\label{subsec:performance_attribution}

To further evaluate the impact of the proposed techniques, we conduct experiments by training the 32B model with the CommonCrawl dataset over 64 GPUs.

\mysubsubsection{Case study}
Figure~\ref{subfig:case_study_1} shows the optimal strategies selected by \system, along with the corresponding latencies, and summarizes the frequency of different strategies over 100 iterations. It can be seen that: (a) the actual propagation latency closely matches our estimation in most cases, and (b) variations across iterations lead to different strategy choices. 

To understand why using different strategy choices contributes, we compare the estimated latencies of two strategies, $S_0$ and $S_3$, over different iterations. 
While $S_3$ achieves $1.6\times$ speedup than $S_0$ in most cases due to its heterogeneous design, its maximum supportable sequence length is shorter than $S_3$, making it inapplicable for certain iterations. 
Besides, when the sampled batch contains fewer short sequences, it's hard for $S_3$ to strike a balance (i.e., many sequences can only be processed by the pipeline with a TP degree of 8 while the other pipelines receive very few sequences), making it less efficient than $S_0$.
These findings highlight the advantages of heterogeneous strategies, as well as the trade-off between latency and maximum sequence length across different strategies, demonstrating the benefits of supporting dynamic heterogeneous strategies for overall optimal performance.

Additionally, we examine the load balancing. Figure~\ref{subfig:case_study_2} shows that Megatron-LM, with max-length packing and even dispatching of packed sequences, results in a latency difference of 0.36-0.65$\times$ between the slowest and fastest ranks. In contrast, \system reduces this gap to just 0.02-0.10$\times$, verifying the significance of the two-stage sequence assignment.

\mysubsubsection{Ablation study}
Figure~\ref{fig:ablation} presents an ablation study by progressively enabling different techniques in \system to evaluate their contributions: (a) Starting with a static, homogeneous parallel strategy, %(i.e., DP, TP, PP),
we dispatch sequences of varying lengths according to Eq.~\eqref{eq:dispatching_ilp} and pack them to the maximum length. 
(b) Instead of packing sequences to the maximum length, we apply packing based on the results of Eq.~\eqref{eq:packing_ilp}. (c) We replace the original homogeneous parallel strategy with a heterogeneous strategy that supports the maximum sequence length. (d) Finally, we enable dynamic heterogeneous strategies, utilizing all \system designs. 

The results demonstrate that, on the one hand, when no optimizations are applied (i.e., Hydraulis-Base), our performance is on par with Megatron-LM, confirming the fairness of the comparison. On the other hand, consistent improvements are observed across (a)-(d), highlighting the critical role each system component plays in boosting overall performance.

\begin{table}[t!]
    \centering
    \small
    \caption{\small{Breakdown of latency per iteration for different training strategies. ``P'' denotes propagation (Figure~\ref{subfig:disaggregate_op} step 3), ``O'' refers to optimization (Figure~\ref{subfig:disaggregate_op} step 6), and ``O-P'' represents the remaining overhead from the interaction between the two phases (Figure~\ref{subfig:disaggregate_op} steps 1-2 and 4-5). ``E2E'' stands for end-to-end training latency.}}
    % \resizebox{\linewidth}{!}{ 
    \begin{tabular}{lccc}
    \hline\toprule
     & \textbf{Homo} & \textbf{Hetero w/o Overlap} & \textbf{Hetero w/ Overlap} \\ \midrule 
    P & 6.78s & 4.29s ($1.58\times$) & 4.29s ($1.58\times$) \\
    O & 0.02s & 0.02s ($1.00\times$) & 0.02s ($1.00\times$) \\
    O-P & 1.11s & 1.71s ($0.69\times$) & 1.18s ($0.94\times$) \\ \midrule
    E2E & 7.91s & 6.02s ($1.31\times$) & 5.49s ($1.44\times$) \\ \bottomrule\hline
    \end{tabular}
    % }
    \label{tab:latency_breakdown}
    \myvspace{-10pt}
\end{table}

\mysubsubsection{Latency breakdown}
We further investigate the effectiveness of optimization-propagation disaggregation compared to conventional model states sharding, as well as the benefits brought by subgraphs scheduling. As shown in Table~\ref{tab:latency_breakdown}, we first adopt a fixed homogeneous parallel strategy (with the two-stage sequence assignment enabled), which utilizes the conventional all-gather and reduce-scatter interaction patterns (Figure~\ref{subfig:standard_op}).
Then, we employ dynamic heterogeneous strategies, introducing additional LNK operators (Figure~\ref{subfig:disaggregate_op}). Besides, we experiment both with and without subgraphs scheduling to measure the latency improvements from overlapping \textit{pull} and \textit{push} interactions with computation.

It can be observed that the use of dynamic heterogeneous strategies improves the performance of propagation. However, the interaction between optimization and propagation becomes slower due to the additional communication overhead introduced by LNK. Despite this, the overhead remains relatively small and can be efficiently reduced by overlapping with computation, leading to only a minimal increase in latency. Moreover, this additional latency does not scale with the number of tokens per iteration, demonstrating the viability of disaggregating optimization and propagation.

%% file: sections/ending.tex
% \myvspace{-8pt}
\section{Related Work}
\label{sec:related_work}

\mysubsubsection{Optimizations of distributed training}
Numerous studies aim to enhance the efficiency of distributed training. In the realm of parallel methods, advancements in pipeline parallel focus on minimizing bubble time through enhanced scheduling~\cite{qi2023zerobubble, lamypoirier2023breadthfirst, jiang2024dynapipe, wu2024bitpipe}, while tensor parallel has progressed by overlapping communication with computation~\cite{wang2022overlap, jangda2022coconet, chen2024centauri, jiang2024megascale}. In the context of model states sharding, efforts have been made to balance memory usage and latency through finer-grained sharding~\cite{chen2024internevo}, efficient communication~\cite{wang2023zero, maurya2024deep}, and multi-level storage solutions~\cite{ren2021zerooffload, memo}. Notably, these approaches are compatible with ours. 
There are also efforts focusing on automatically searching for the optimal parallel strategies~\cite{zheng2022alpa, jia2019beyond, miao2022galvatron, wang2024improving, hetu-v2, spindle}. 
However, they only consider homogeneous strategies and overlook the data imbalance issues.

\mysubsubsection{Resources disaggregation}
The idea of disaggregating storage from compute resources has been well-studied in cloud computing~\cite{shan2018legoos, guo2023mira, lim2012system} and utilized in many AI applications~\cite{audibert2023tfdata, jin2024distmind}. Similar concepts have also gained attention in large Transformer model inference, where disaggregating the prefilling and decoding phases allows for better throughput~\cite{patel2024splitwise, hu2024inference, zhong2024distserve}. In our system, we adopt a similar design philosophy by separating the storage-heavy optimization phase from the compute-intensive propagation phase, enabling flexible execution and transitions of parallel strategies. 

\mysubsubsection{Intermediate representations of parallel strategies}
Graph-based intermediate representations are widely utilized in deep learning systems~\cite{abadi2016tensorflow, lattner2021mlir, jin2020compilingonnx, hagedorn2023graphene}. By analyzing and optimizing the graphs, models can discover enhanced and specialized parallel strategies. \system extends this by simultaneously deducing and expressing graphs for different strategies. Using subgraphs, it distinctly highlights the unique communication patterns of each strategy. Besides, any graph-based optimization techniques can be applied to \system, allowing for further refinement for each strategy.

\section{Conclusion and Future Work}
\label{sec:conclusion}

In this work, we addressed the often-overlooked issues of data sampling and packing imbalances in existing training systems, and introduced \system, a novel system that leverages parallel computing and data management co-designs to optimize the training performance of large Transformer models. 
We proposed training with dynamic heterogeneous parallel strategies to address the data sampling imbalance and a two-stage sequence assignment to tackle the data packing imbalance. 
Empirical results show that \system can achieve up to a $2.66\times$ increase in throughput compared to state-of-the-art training systems. 

As a possible future direction, we intend to expand our strategy space by integrating more parallel methods (e.g., expert parallel~\cite{GShard, netmoe}), as our system designs are fully compatible with more forms of parallelism. We also plan to tackle data imbalance challenges in diverse domains (e.g., in the video domain, the significant variation in sequence lengths has already attracted considerable attention~\cite{chen2024timemarkerversatilevideollmlong, liu2024ppllavavariedvideosequence, shang2024interpolatingvideollmslongersequencelmms, adaspa}). We believe our work can be adapted to a broader range of scenarios and applications.

%% file: sections/appendix.tex
\section{Data Imbalance}
\label{appendix:imbalance}

In this section, we will provide more evidence to support claims about data imbalance in \S\ref{sec:analysis}. Firstly, we show that increasing the batch size does not really eliminate the variation in maximum sequence length across iterations. Subsequently, we conduct experiments to examine whether the quadratic time complexity of the attention mechanism impacts end-to-end performance.

\begin{figure}[!t]
    \centering
    \includegraphics[width=\linewidth]{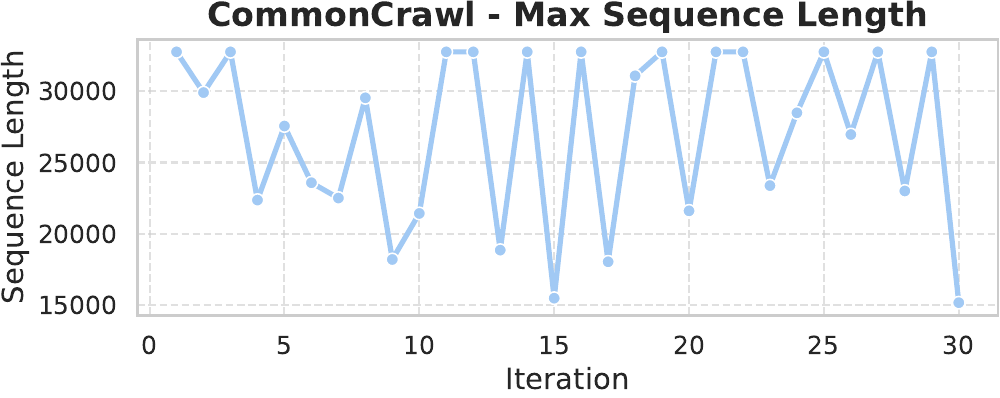}
    \caption{\small{Curve of maximum sequence length across iterations. We sample 1.6M tokens from the CommonCrawl dataset per iteration and track the longest sequence.}}
    \label{fig:larger_batch}
\end{figure}

\subsection{Larger Batch Size}
\label{appendix:larger_batch_size}

In our experiment, we use 64 GPUs, processing 100K tokens per iteration. 
Intuitively speaking, as we scale up to 1024 GPUs, the number of tokens per iteration should increase to 1.6M. 
To examine whether a larger batch size would resolve the inter-iteration imbalance, we further conduct an experiment to sample 30 consecutive iterations on the CommonCrawl dataset, with 1.6M tokens per iteration.
Throughout this process, we observe that the maximum sequence length still fluctuates, and falls between 15K and 32K. While this fluctuation is less pronounced compared to the scenario of using 100K tokens per iteration, it still indicates an apparent inter-iteration imbalance.

\subsection{End-to-end Influence of Attention}
\label{appendix:varlen_attn}

\begin{figure}[!t]
    \centering
    \includegraphics[width=\linewidth]{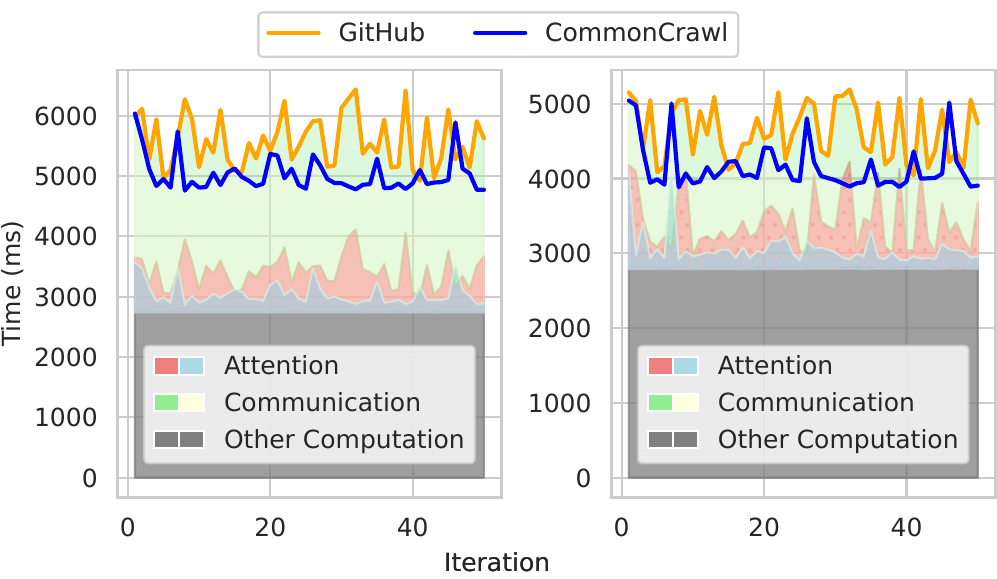}
    \caption{\small{Latency breakdown and fluctuation when training LLaMA 7B on 8 Nvidia A800 GPUs. The left uses pipeline parallel (PP) without data parallel (DP), while the right uses data parallel (DP) without pipeline parallel (PP).}} 
    \label{fig:attn_imbalance}
\end{figure}

To better demonstrate the impact of the quadratic time complexity of the attention mechanism on end-to-end training, we train the 7B model. Sequences are packed in mini-batches using the first-fit-decreasing bin packing algorithm~\cite{nvidia2023packing}, with a maximum length of 32K. Each mini-batch contains 100K tokens, and we manually ensure that they will form 4 packed sequences (if not, we just skip that iteration for this demonstration experiment). Figure~\ref{fig:attn_imbalance} shows the results. Despite carefully controlling tokens and the number of packed sequences, training time still fluctuates by about 20\% in both settings. This variation primarily stems from differences in the computational cost of the attention mechanism, while other computational operations and communication overhead remain relatively stable.

\section{Communication in O-P Disaggregation}
\label{appendix:comm}

\subsection{Communication Bound}
\label{appendix:comm_bound}

In this section, we provide formal proof of the communication bound in O-P disaggregation. 

We consider $N$ GPUs and a model parameter of size $W$, and analyze the communication volume when each GPU participates as either a receiver or a sender during the \textit{pull} operation, denoted as $V_{\text{recv}}$ and $V_{\text{send}}$, respectively.

For the receiver side analysis, consider a GPU belonging to a pipeline that adopts a parallel scheme $P=\langle TP, PP, CP \rangle$ during the propagation phase. The GPU must receive the complete parameter partition corresponding to its pipeline stage. Specifically:
\begin{itemize}[leftmargin=*]
    \item If the parameter resides in the GPU's current pipeline stage, it needs to receive a parameter partition of size $\frac{W}{TP}$.
    \item If the parameter does not belong to the current stage, the receiving volume becomes $0$ as no pull operation is required.
\end{itemize}
Therefore, we have:
\[
V_{\text{recv}} = \frac{W}{TP} \quad \text{or} \quad V_{\text{recv}} = 0
\]

For the sender side analysis, our fully-sharded partitioning strategy during the optimization phase ensures each GPU stores exactly one parameter slice of size $\frac{W}{N}$. When $M$ GPUs (where $M \leq N$) request this particular parameter slice during propagation, the sending GPU's communication volume is given by:
\[
V_{\text{send}} = \frac{W}{N} \times M
\]

Based on this, we can directly derive the worst-case communication scenario, which occurs when the parallel strategy takes the form  $S=\langle 1,1,C \rangle \times \frac{N}{C}$, where $C$ is any positive integer divisor of $N$. This configuration represents the case where:
\begin{itemize}[leftmargin=*]
    \item The system contains no tensor or pipeline parallelism ($TP=1$, $PP=1$).
    \item Every GPU requires all parameter slices ($M=N$).
\end{itemize}
Consequently, each GPU's communication volume reaches its theoretical maximum:
\[
V_{\text{recv}} = \frac{W}{1} = W \quad \text{and} \quad V_{\text{send}} = \frac{W}{N} \times N = W
\]
This worst-case scenario exactly corresponds to training with a homogeneous fully-sharded ZeRO-1 strategy, where the communication cost per GPU is bounded by $W$ and becomes independent of the total GPU count $N$. 

An analogous analysis applies to the \textit{push} operation.

\subsection{Computation-Communication Overlap}
\label{appendix:comm_overlap}

In this section, we analyze the condition for overlapping communication with computation based on the derived communication bound.

For a matrix multiplication operation under the pipeline with parallel scheme $\langle \mathit{TP}, \mathit{PP}, \mathit{CP} \rangle$, consider input $X$ with shape $[t/\mathit{CP}, h]$ and weight matrix $W$ with shape $[h, h'/\mathit{TP}]$, where $t$ denotes the number of tokens, $h$ and $h'$ represent hidden state dimensions. During the computation of this matrix
multiplication, we aim to \textit{pull} the parameter for the next matrix
multiplication. As established in Appendix~\ref{appendix:comm_bound}, the maximum amount
of data communicated would be $h \times h'$ (i.e., the whole parameter size). Assuming half-precision
(e.g., BF16) is used, the total  communication volume is:
\[
\text{Comm}_{\text{bytes}} = 2hh'
\]
And the computational workload (FLOPs) for the matrix multiply (counting multiply-add as 2 operations) is:
\[
\text{FLOPs} = 2 \cdot \underbrace{\left(\frac{t}{\mathit{CP}}\right)}_{\text{rows}} \cdot \underbrace{h}_{\text{mid-dim}} \cdot \underbrace{\left(\frac{h'}{\mathit{TP}}\right)}_{\text{cols}} = \frac{2thh'}{\mathit{CP} \cdot \mathit{TP}}
\]

Let $F$ denote the GPU's half-precision compute capability (FLOPS) and $B$ the interconnect bandwidth (bytes/sec). The computation time $T_{\text{comp}}$ and communication time $T_{\text{comm}}$ are:
\begin{align*}
T_{\text{comp}} & = \frac{\text{FLOPs}}{F} = \frac{2thh'}{\mathit{CP} \cdot \mathit{TP} \cdot F} \\ 
T_{\text{comm}} & = \frac{\text{Comm}_{\text{bytes}}}{B} = \frac{2hh'}{B}
\end{align*}
To overlap communication with computation, we require:
\[
T_{\text{comp}} \geq T_{\text{comm}}
\]
Substituting the expressions and simplifying:
\[
\frac{2thh'}{\mathit{CP} \cdot \mathit{TP} \cdot F} \geq \frac{2hh'}{B}
\]
Canceling common terms ($2hh' \neq 0$):
\[
\frac{t}{\mathit{CP} \cdot \mathit{TP} \cdot F} \geq \frac{1}{B}
\]
Solving for $t$:
\[
t \geq \mathit{TP} \cdot \mathit{CP} \cdot \frac{F}{B}
\]
This indicates that overlapping requires the token count $t$ to exceed a threshold determined solely by the parallel scheme ($\mathit{TP}$, $\mathit{CP}$) and hardware ratio ($F/B$), independent of parameter dimensions $h$, $h'$.

As a concrete example, consider an A800 cluster running the 32B model with a pipeline configured with $\left<16,1,1\right>$, corresponding to Figure~\ref{subfig:case_study_1}. The hardware specifications are (assuming
the worst-case scenario where all communication goes through
InfiniBand):
\begin{align*}
F & = 19.5 \times 10^{12}  \text{ FLOPS} \\ B & = 200 \times 10^9  \text{ bytes/sec (worst-case)}
\end{align*}
The token threshold is then:
\begin{align*}
t_{\text{threshold}} & = \mathit{TP} \cdot \mathit{CP} \cdot \frac{F}{B} = 16 \times 1 \times \frac{19.5 \times 10^{12}}{200 \times 10^9} = 1560
\end{align*}
This means that, theoretically, sequence lengths exceeding 1.5K after packing are sufficient to enable overlap (note that both GPU compute power and communication bandwidth may vary in real-world settings, so adjustments will be made based on empirical measurements). In practice, this condition is fully achievable in our setup.

\section{Cost Model}
\label{appendix:cost_model}

In this section, we elaborate on the memory model $\texttt{MaxLen}(P)$, which determines the maximum sequence length, and the latency model $\texttt{T}(l, P)$, which estimates the propagation time for a specific sequence length $l$, given a parallel scheme $P$.

\subsection{Memory Model}
\label{appendix:memory_model}
Consider a model with hidden size $H$, number of layers $\overline{\textup{layers}}$, and vocab size $V$, and there are $N$ available GPUs. During the propagation phase, we utilize the parallel scheme $\left<TP, PP, CP\right>$, while the optimization phase employs a fully sharded strategy. For a sequence length $l$ (i.e., micro-batch shape $[1, l]$), memory usage is calculated as follows. (a) For activations, the memory cost per micro-batch is $\frac{\overline{\textup{layers}}}{PP} \times \frac{l \times H}{TP \times CP} \times A$, where $A$ is a constant related to the model architecture, determined through profiling. The cumulative memory peak of all micro-batches (accumulated $PP$ times in the first pipeline) is $\overline{\textup{layers}} \times \frac{l \times H}{TP \times CP} \times A$. (b) For model states, given that some are fully sharded while others are related to $\left<TP, PP, CP\right>$, the total memory usage of all layers is $\frac{\overline{\textup{layers}} \times H^2}{N}  \times B \times \alpha + \frac{\overline{\textup{layers}}}{PP} \times \frac{H^2}{TP} \times B \times \left(1 - \alpha\right)$. Here, $B$ is a constant related to the model structure, obtainable via profiling, and $\alpha$ is associated with the distribution and precision of model states. For instance, if the optimization phase uses 32-bit optimizer states and parameters, while the propagation phase uses 16-bit parameters and gradients, then $\alpha=\frac{32 \times 3}{32 \times 3 + 16 \times 2}=\frac{3}{4}$. Similarly, for the embedding table, the memory cost is $\frac{H \times V}{N} \times 16 \times \alpha + \frac{H \times V}{TP} \times 16 \times \left(1 - \alpha\right)$. The total memory peak is the sum of these costs, forming a linear function of $l$. Then, the x-coordinate of this function at the y-coordinate corresponding to GPU memory capacity minus a safety margin provides the maximum sequence length $\texttt{MaxLen}(P)$ supported by $P = \left<TP, PP, CP\right>$. It’s important to note that we can apply a similar modeling approach when incorporating other forms of parallel methods, such as context parallel and sequence parallel.

\subsection{Latency Model}
\label{appendix:latency_model}

The latency is directly modeled as $\texttt{T}\left(l, P_k\right) = a_k \times l^2 + b_k \times l + c_k$, where $a_k, b_k, c_k$ are coefficients related to $P_k$. The quadratic term reflects the quadratic scaling of attention computation with sequence length, while the linear term accounts for other computations and activation communications. The constant term represents additional overheads such as kernel launch delays. By profiling for $P_k$ and adjusting $l$, we fit the curve and determine these coefficients.

\section{Enumeration Pruning}
\label{appendix:enumeration_pruning}

\begin{figure}[!t]
    \begin{subfigure}{0.59\linewidth}
        \centering
        \includegraphics[width=1\linewidth]{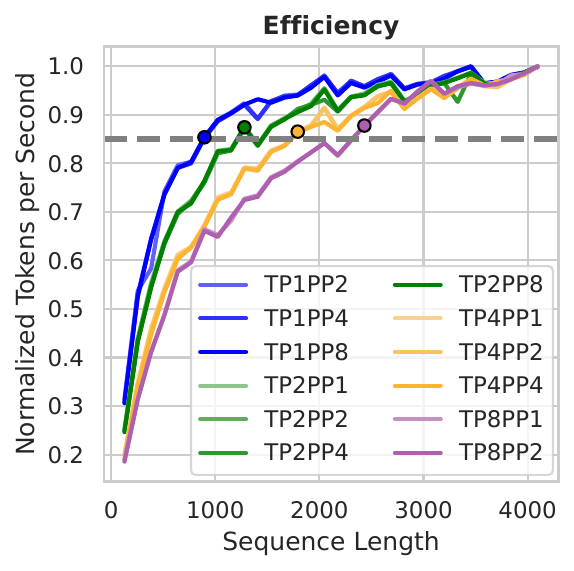}
        \subcaption{Efficiency comparison.}
        \label{subfig:gpu_utilization}
    \end{subfigure}
    \begin{subfigure}{0.4\linewidth}
        \centering
        \includegraphics[width=1\linewidth]{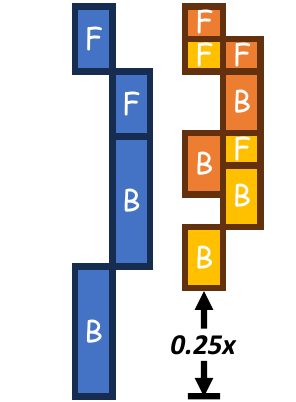}
        \subcaption{Bubble comparison.}
        \label{subfig:bubble}
    \end{subfigure}
    \caption{\small{(a) LLaMA 13B training efficiency across different parallel schemes, measured in tokens per second. (b) Latency changes pre- and post-sequence packing for two equal-length sequences when $PP=2$, ``F'' and ``B'' indicate forward and backward propagation, respectively.}}
    \label{fig:strategies_property}
\end{figure}

In this section, we will further discuss two trade-offs during packing as a supplement. Based on this, we can perform pruning optimization to avoid some unnecessary enumeration of the number of micro-batches.

\begin{enumerate}[leftmargin=*]
    \item \textbf{Less packing, smaller micro-batch size, lower GPU utilization:} As shown in Figure~\ref{subfig:gpu_utilization}, we assess the GPU's efficiency in processing sequences of specific lengths by dividing the propagation latency by the number of tokens in the sequence. Our observations indicate that efficiency approaches optimal only when the sequence length exceeds a certain threshold, which is related to intra-op parallel (e.g., TP) but independent of inter-op parallel (e.g., PP). This is due to GEMMs (General Matrix Multiplications)~\cite{nvidia2023matrixmultiplication}: intra-op parallel reduces the computational load of single matrix multiplication, while inter-op parallel only reduces the number of matrix multiplications. When sequences are short or the intra-op parallel degree is large, the scale of a single matrix multiplication becomes small, leading to under-utilization of the GPU's SMs and significant kernel launch overhead~\cite{nvidia2023gpuperformance}. Therefore, insufficient packing on some parallel schemes may result in many under-utilized micro-batches, reducing the performance.
    \item \textbf{More packing, smaller number of micro-batches, higher pipeline bubble rate:} As shown in Figure~\ref{subfig:bubble}, packing inevitably reduces the number of micro-batches, increasing the bubble rate. In the example shown, there is about a $0.25\times$ performance slowdown. Therefore, packing sequences to the memory limit or the mini-batch maximum is not necessarily optimal.
\end{enumerate}

Based on these two observations, we propose that the number of micro-batches should be neither too large nor too small. Otherwise, it could lead to low GPU utilization or a high bubble rate, respectively. Therefore, we first profile each parallel scheme and introduce the value $\texttt{UtilLen}\left(P\right)$, representing the fully utilized sequence length of $P$ at which its efficiency curve reaches a certain threshold (0.85 in our experiments). This provides an upper bound for the number of micro-batches. Additionally, the maximum sequence length supported by the pipeline configured with $P$ (i.e., $\texttt{MaxLen}\left(P\right)$) offers a good lower bound. Denote the sequences dispatched to the pipeline with parallel scheme $P$ as $l_1, \ldots, l_U$ and $V$ as the number of micro-batches, then the enumeration of $V$ can be pruned from the original $V \in [1, U]$ to:
\begin{equation*}
V \in \left[\max\left(\frac{\sum\limits_{i=1}^{U} l_i}{\texttt{MaxLen}(P)}, 1\right), \min\left(\frac{\sum\limits_{i=1}^{U} l_i}{\texttt{UtilLen}(P)}, U\right)\right]
\end{equation*}

\section{Details of Sequence Dispatching across Pipelines}
\label{appendix:dispatching_datails}

In this section, we begin by deriving Eq.~\eqref{eq:dispatching_lower_bound}. We then describe the approximation algorithm for solving the sequence dispatching problem across pipelines (i.e., Eq.~\eqref{eq:dispatching_ilp}) and compare it to the precise solution.

\subsection{Derivation of the Lower Bound}
\label{appendix:derivation_of_lower_bound}

\begin{algorithm}[!t]
\caption{Greedy-based approximate solver for sequence dispatching across pipelines.}
\label{alg:greedy-dispatching}
\KwIn{
    Sequence lengths $\{l_1, \ldots, l_B\}$. Maximum strategy indices $\{J_1, \ldots, J_B\}$. Parallel schemes of the pipelines $\left\{{P}_1, \ldots, {P}_D\right\}$. Number of trials $T$\;
}
\KwOut{
    Optimal cost $O_\text{best}$. Dispatch matrix $M_\text{best}$\;
}

Initialize $O_{\text{best}} \leftarrow \infty$, $M_{\text{best}} \leftarrow \emptyset$\;

\ForPar{trial $= 1$ to $T$}{
    Randomize permutation $\pi$\ to shuffle sequences\;
    Initialize $C_j, E_j, m_{i,j} \leftarrow 0$ , $\forall i \in [1, B], j \in [1, D]$\;

    \For{$k = 1$ to $B$}{
        $i \leftarrow \pi_k$, $O_{\text{min}} \leftarrow \infty$, $j^* \leftarrow -1$\;

        \For{$j = 1$ to $J_i$}{
            $l_\text{max} \leftarrow \max(l_j, \max_{i}\{m_{i,j} \times l_i\})$\;
            $C_j' \leftarrow C_j + \texttt{T}\left(l_i, {P}_j\right)$\;
            $E_j' \leftarrow \texttt{T}\left(l_\text{max}, {P}_j\right) \times \left(\texttt{PP}\left(P_j\right) - 1\right)$\;
            $O_{\text{max}} \leftarrow \max(C_j' + E_j', C_k + E_k\text{ for } k \neq j)$\;

            \If{$O_{\text{max}} < O_{\text{min}}$}{
                $O_{\text{min}} \leftarrow O_{\text{max}}$, $j^* \leftarrow j$\;
            }
        }
        $m_{i,j^*} \leftarrow 1$, $C_{j^*} \leftarrow C_{j^*}'$, $E_{j^*} \leftarrow E_{j^*}'$\;
    }

    $O_{\text{trial}} \leftarrow \max_{j}\{C_j + E_j\}$\;

    \If{$O_{\text{trial}} < O_{\text{best}}$}{
        $O_{\text{best}} \leftarrow O_{\text{trial}}$, $M_{\text{best}} \leftarrow \{m_{i,j}\}_{B\times D}$\;
    }
}
\Return $O_{\text{best}}$, $M_{\text{best}}$\;
\end{algorithm}

Consider Eq.~\eqref{eq:packing_ilp} for the $d$-th pipeline. Let the optimal solution be $o^*_{i,j}$, then the optimal objective can be split into two terms:
\begin{equation*}
\small
\begin{aligned}
A & = \max\limits_{1 \le j \le V}{\left(\sum\limits_{1 \le i \le U}{o^*_{i,j} \times \texttt{T}\left(l_i, P_d\right)}\right) \times (\texttt{PP}(P_d) - 1)} \\
B & = \max\limits_{1 \le j \le V}{\left(\sum\limits_{1 \le i \le U}{o^*_{i,j} \times \texttt{T}\left(l_i, P_d\right)}\right) \times V} \\
\end{aligned}
\end{equation*}
For $A$, consider the longest dispatched sequence length, denoted as $l_{i_{\star}} = \max_{1 \le i \le B}\left\{m_{i,d} \times l_i\right\}$. Since $\sum_{1 \le j \le V}{o^*_{i_{\star},j}} = 1$ and $o^*_{{\star},j} \in \left\{0, 1\right\}$, there must exist a $j_{\star}$ such that $o^{*}_{i_\star,j_\star} = 1$. Thus, we have:
\begin{equation*}
\small
\begin{aligned}
A & \ge \sum\limits_{1 \le i \le U}{o^*_{i,j_\star} \times \texttt{T}\left(l_i, P_d\right)} \times (\texttt{PP}(P_d) - 1) \\
& \ge o^*_{i_\star,j_\star} \times \texttt{T}\left(l_{i_\star}, P_d\right) \times (\texttt{PP}(P_d) - 1) \\ 
& \ge \texttt{T}\left(\max_{1 \le i \le B}\left\{m_{i,j} \times l_i\right\}, P_d\right) \times (\texttt{PP}(P_d) - 1)
\end{aligned}
\end{equation*}
For $B$, it is evident that:
\begin{equation*}
\small
\begin{aligned}
\sum\limits_{1 \le i \le U}{o^*_{i,j} \times \texttt{T}\left(l_i, P_d\right)} &\le \max\limits_{1 \le j \le V}{\left(\sum\limits_{1 \le i \le U}{o^*_{i,j} \times \texttt{T}\left(l_i, P_d\right)}\right)}, \forall j \in \left[1, V\right]
\\ 
\implies B &= \max\limits_{1 \le j \le V}{\left(\sum\limits_{1 \le i \le U}{o^*_{i,j} \times \texttt{T}\left(l_i, P_d\right)}\right) \times V} \\ 
& \ge \sum\limits_{1 \le j \le V}{\left(\sum\limits_{1 \le i \le U}{o^*_{i,j} \times \texttt{T}\left(l_i, P_d\right)}\right)} \\
& = \sum\limits_{1 \le i \le B} m_{i,d} \times \texttt{T}(l_i, P_d)
\end{aligned}
\end{equation*}
By summing these two terms, the dispatching lower bound for the $d$-th pipeline is obtained:
\begin{equation*}
\small
\sum_{i=1}^B m_{i,d} \times \texttt{T}(l_i, P_d) + \texttt{T}\left(\max_{1 \le i \le B}\left\{m_{i,d} \times l_i\right\}, P_d\right) \times \left(\texttt{PP}\left(P_d\right) - 1\right) 
\end{equation*}

\subsection{Approximate Solver}

Our developed approximation algorithm is presented in Alg.~\ref{alg:greedy-dispatching}. We start by randomly shuffling the sequences and then dispatch them one by one (lines 1-6). For each dispatch, we aim to prevent the current slowest pipeline $P_j$ from worsening, by considering the accumulated sequence cost $C_j$ and extra cost $E_j$ (lines 7-14). Specifically, for each potential pipeline $j$ of the current sequence $i$ (line 7), we simulate dispatching the sequence to observe the cost change. We first compute the maximum sequence length based on the current sequence and those already dispatched to the pipeline (line 8). Then, we calculate the new accumulated cost $C_j^\prime$ by adding the latency of the current sequence $i$ on pipeline $j$ (line 9). After determining the new extra cost $E_j^\prime$, which is associated with the maximum sequence length and the pipeline parallel (PP) degree (line 10), we evaluate the maximum total cost $O_\text{max}$ across all pipelines (line 11). Afterward, we select the pipeline $j^*$ that minimizes $O_\text{max}$ and update the dispatch matrix and costs accordingly (lines 12-14). This process (i.e., a trial) is repeated $T$ times to select the best result (lines 15-17).\footnote{$T$ is set to 100 in all our experiments.} Since multiple trials can be parallelized using multiple processes, the actual execution speed is relatively fast (details in Appendix~\ref{appendix:planning_time}), and the gap compared to the optimal solution of Eq.~\eqref{eq:dispatching_ilp} is relatively small (details in Appendix~\ref{appendix:dispatching_compare}).

\subsection{Performance Comparison}
\label{appendix:dispatching_compare}

\begin{figure}[t!]
    \centering
    \includegraphics[width=\linewidth]{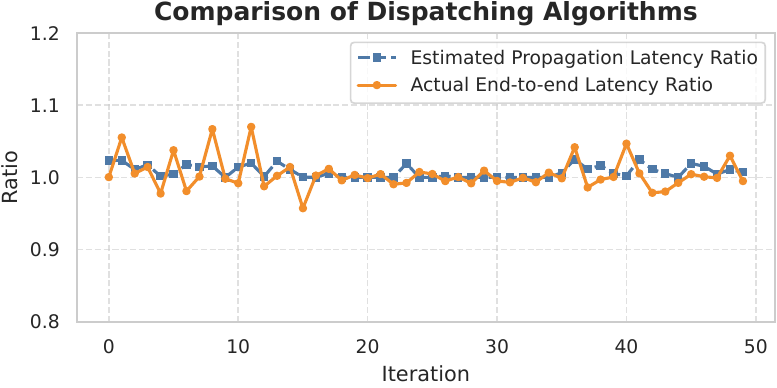}
    \caption{\small{Comparison of estimated propagation latency and actual end-to-end latency between \textit{greedy dispatching} and \textit{ideal dispatching}. The experiments use a 32B model on the CommonCrawl dataset, with 64 GPUs. Each iteration handles 200,000 tokens with a maximum sequence length of 32K.}}
    \label{fig:dispatch_comparison}
\end{figure}

\begin{table*}[!t]
    \centering
    \caption{\small{Solving time for each subproblem. The time cost for \textit{ideal dispatching}, \textit{greedy dispatching}, and \textit{packing} is listed for \textbf{10 iterations} combined. \system ultimately chooses the \textit{greedy dispatching} scheme over the \textit{ideal dispatching} scheme due to its faster speed.}}
    \small
    \begin{tabular}{lcccc}
    \toprule
    & \textbf{Strategy Proposal} & \textbf{Ideal Dispatching} & \textbf{Greedy Dispatching} & \textbf{Packing} \\
    \midrule
    64 GPUs \& 200K tokens per iteration & 2.03s  & 40.21s   & 5.46s   & 10.15s \\
    1024 GPUs \& 3.2M tokens per iteration & 107.21s & >1000s   & 38.56s  & 13.46s \\
    \bottomrule
    \end{tabular}
    \label{tab:planning_time}
\end{table*}

\begin{table*}[t!]
    \centering
    \caption{\small Optimal training configurations. DP refers to the data parallel degree, TP to the tensor parallel degree, PP to the pipeline parallel degree, CP to the context parallel degree, SP to the Ulysses sequence parallel degree, and AC indicates the use of activation checkpointing. For HotSPa, there are 2-3 distinct strategies.}
    \small
    \renewcommand{\arraystretch}{1.5} % 增加行高
    \resizebox{\textwidth}{!}{
    \begin{tabular}{lcccc}
    \hline\toprule
    \multirow{2.25}{*}{\textbf{LLaMA 7B}} & \multicolumn{2}{c}{\textbf{8 GPUs}} & \multicolumn{2}{c}{\textbf{16 GPUs}} \\
    \cmidrule(r){2-3} \cmidrule(l){4-5}
    & F1 & F2 \& F3 & F1 & F2 \& F3 \\
    \midrule
    DeepSpeed      & DP2SP4     & SP8               & DP4SP4     & DP2SP8 \\
    Megatron-LM    & DP2TP2PP2  & TP4PP2           & DP4TP2PP2  & DP2TP4PP2 \\
    HotSPa         & DP2TP2PP2, DP8 & TP4PP2, DP4TP2PP2, DP2TP4PP2 & DP4TP2PP2, DP16 & DP2TP4PP2, DP4TP2PP2, DP16 \\
    \midrule
    \multirow{2.25}{*}{\textbf{LLaMA 13B}} & \multicolumn{2}{c}{\textbf{16 GPUs}} & \multicolumn{2}{c}{\textbf{32 GPUs}} \\
    \cmidrule(r){2-3} \cmidrule(l){4-5}
    & F1 & F2 \& F3 & F1 & F2 \& F3 \\
    \midrule
    DeepSpeed      & DP4SP4     & DP2SP8           & DP8SP4     & DP4SP8 \\
    Megatron-LM    & DP4TP4     & DP2TP8           & DP8TP4     & DP4TP8 \\
    HotSPa         & DP4TP4, DP8TP2 & DP2TP8, DP4TP4, DP8TP2 & DP8TP4, DP16TP2 & DP4TP8, DP8TP4, DP16TP2 \\
    \midrule
    \multirow{2.25}{*}{\textbf{LLaMA 32B}} & \multicolumn{2}{c}{\textbf{32 GPUs}} & \multicolumn{2}{c}{\textbf{64 GPUs}} \\
    \cmidrule(r){2-3} \cmidrule(l){4-5}
    & F1 & F2 \& F3 & F1 & F2 \& F3 \\
    \midrule
    DeepSpeed      & DP8SP4 + AC & DP4SP8 + AC      & DP16SP4 + AC & DP8SP8 + AC \\
    Megatron-LM    & DP2TP8PP2   & DP2TP8CP2          & DP4TP8PP2   & DP4TP8CP2 \\
    HotSPa         & DP2TP8PP2, DP4TP4PP2 & DP2TP16, DP2TP8PP2, DP4TP4PP2 & DP4TP8PP2, DP8TP4PP2 & DP4TP16, DP4TP8PP2, DP8TP4PP2 \\
    \bottomrule\hline
    \end{tabular}
    }
    \label{tb:config}
\end{table*}

\begin{table*}[!t]
    \centering
    \caption{\small{HotSPa's switching overhead and iteration time comparison.}}
    \resizebox{\textwidth}{!}{
    \begin{tabular}{lcccccc}
    \toprule
    Configuration & \textbf{Tokens per Iteration} & \textbf{HotSPa Switching Time} & \textbf{HotSPa Iteration Time (incl. Switching)} & \textbf{Hydraulis Iteration Time} \\
    \midrule
    (a) 7B, 32 GPUs & 100 & 1.54s & 5.86s & 2.21s \\
    (b) 32B, 32 GPUs & 100 & 2.17s & 12.21s & 7.21s \\
    (c) 32B, 64 GPUs & 200 & 2.21s & 13.19s & 7.22s \\
    \bottomrule
    \end{tabular}}
    \label{tab:hotspa}
\end{table*}

We analyze the gap between our proposed dispatching algorithm, Alg.~\ref{alg:greedy-dispatching} (referred to as \textit{greedy dispatching}), and the optimal solution to the original optimization problem, Eq.~\eqref{eq:dispatching_ilp} (referred to as \textit{ideal dispatching}).
As illustrated in Figure~\ref{fig:dispatch_comparison}, we compare the ratios of estimated propagation latency (i.e., the optimal objective given by the algorithm) and actual end-to-end latency by dividing the \textit{greedy dispatching} results by the \textit{ideal dispatching} results. The estimations closely align, and the actual end-to-end latency ratios fall between 0.9 and 1.1. This indicates that there is no significant performance difference between the two approaches.
Considering the subsequent time analysis in Appendix~\ref{appendix:planning_time}, we ultimately opt for \textit{greedy dispatching} due to its faster solution speed and superior scalability.

\section{Time Cost and Scalability}
\label{appendix:planning_time}

In this section, we discuss the computation times of all mathematical models and their scalability. The experiments are conducted on the CommonCrawl dataset using a 32B model configuration with 64 GPUs, a maximum sequence length of 32K, and 200K tokens per iteration.

Table~\ref{tab:planning_time} presents the solving time for various components of our mathematical model. To enhance computational efficiency, we employ multi-processing for all components except the \textit{strategy proposal}. The \textit{strategy proposal} process relies on a dynamic programming algorithm, which is inherently sequential and, therefore, not suitable for parallelization. For the parallelized components, we utilize 10 CPU processes, concurrently solving the corresponding problem for \textbf{10 iterations}, and record the resulting latency. To further study the scalability of our planning algorithms, apart from running real end-to-end training on the existing testbed, we also simulate a scaled-up scenario with 1024 GPUs, where the number of tokens per iteration is proportionally increased to 3.2 million.

For the \textit{strategy proposal}, its computational complexity scales linearly with the number of GPUs in the cluster. Consequently, the time cost increases significantly from 64 GPUs to 1024 GPUs. However, since this component only executes once, its time cost--- remaining under 2 minutes--- is acceptable.

For the \textit{ideal dispatching} (i.e., Eq.~\eqref{eq:dispatching_ilp}), while the time cost is acceptable at 64 GPUs, scaling to 1024 GPUs results in superlinear growth. This is due to the complexity being proportional to both the number of pipelines and the number of sequences per iteration--- both of which grow dramatically at larger scales. In contrast, \textit{greedy dispatching} (i.e., Alg.~\ref{alg:greedy-dispatching}) offers a significantly faster alternative, effectively controlling the growth in time cost. This makes it a more practical choice for large-scale deployments.

For the \textit{packing} (i.e., Eq.~\eqref{eq:packing_ilp}) part, we measure the time required to solve all the ILPs. Each ILP addresses the allocation of sequences within a single pipeline, and its scale remains independent of both the number of GPUs and the total number of sequences per iteration. Moreover, since these ILPs are completely independent of each other, they can be solved in parallel, resulting in a relatively stable solving time across different scales.

Overall, considering the training latency of \system in our experiments, the time 
cost for a single iteration is approximately 5.5 seconds, leading to a total of around 55 seconds for 10 iterations. Regardless of whether the cluster consists of 64 GPUs or 1024 GPUs, the computation time of our algorithms can be effectively overlapped with GPU execution. In the implementation of \system, we employ a dedicated process for training and adopt a producer-consumer model to coordinate planning and training. Specifically, the planning processes, acting as producers, solve the sequence dispatching and packing problems, generating execution plans. Meanwhile, the training process, acting as the consumer, retrieves and executes the plans for the current iteration. This design ensures that GPUs remain fully saturated with training workloads throughout execution, thereby achieving efficient utilization of computational resources.

\section{Optimal Training Configurations}
\label{appendix:configuration}

Table~\ref{tb:config} illustrates the manually tuned configurations that achieve the best training performance under different scenarios in \S\ref{subsec:e2e_exp}. In terms of HotSPa:
\begin{itemize}[leftmargin=*]
    \item For scenario F1, with a maximum sequence length of 16K, two strategies are employed: (a) The first strategy handles sequences of 4K to 16K. (b) The second strategy handles sequences less than 4K.
    \item For scenarios F2 and F3, with a maximum sequence length of 32K, three strategies are employed: (a) The first strategy handles sequences of 16K to 32K. (b) The second strategy handles sequences of 4K to 16K. (c) The third strategy handles sequences less than 4K.
\end{itemize}

\section{HotSPa's Switching Overhead}
\label{appendix:hotspa}

In this section, we analyze the switching overhead of HotSPa under three configurations, as detailed in Table~\ref{tab:hotspa}. From configuration (a) to (b), we scale the model size, and from (b) to (c), we scale the total training setup, proportionally to LLaMA’s configuration (16M tokens per iteration for 8192 GPUs).

First, when the training workload is relatively small, HotSPa’s switching overhead becomes significant. For example, in configuration (a), 1.54s out of the total 5.86s per iteration (26\%) is spent on strategy switching, negatively impacting end-to-end performance.

Second, when scaling from configuration (b) to (c), doubling the data parallel (DP) degrees causes HotSPa’s switching to occur independently within each 32-GPU group. This keeps the switching overhead relatively stable. However, since the tokens per iteration also scale, the overhead remains consistently around 17\% of the total iteration time. (Note that our experimental environment is equipped with 400 GB/s NVLink and 200 GB/s IB, which are sufficiently fast.) In contrast, Hydraulis overlaps strategy switching with computation via O-P disaggregation (as evaluated in \S\ref{sec:exp}, Table~\ref{tab:latency_breakdown}), making it a more scalable solution.

Finally, even if the switching overhead were excluded, HotSPa would still lag behind Hydraulis. This is due to the advantages of Hydraulis in terms of strategy optimization and sequence assignment: For one thing, HotSPa only supports homogeneous strategies at any given time, whereas Hydraulis enables dynamic heterogeneous strategies. For another, Hydraulis incorporates fine-grained sequence dispatching and packing, which HotSPa lacks.